\documentclass[double]{article}
\usepackage{times}
\usepackage{algorithm}
\usepackage{algorithmic}
\usepackage{wrapfig}
\usepackage{verbatim}
\usepackage{amsmath}
\usepackage{graphicx}
\usepackage{subcaption}
\usepackage[active]{srcltx}
\usepackage{color}
\usepackage{lineno}
\usepackage{dirtytalk}
\usepackage{amsthm}
\usepackage{booktabs}
\usepackage{authblk}
\usepackage{listings}
\usepackage{tikz}
\usepackage{longtable}
\usetikzlibrary{shapes.geometric, arrows}

\usepackage{epsfig,graphicx,amsfonts}
\usepackage{amsmath,amsthm,amssymb}
\usepackage{xcolor}

\usepackage{adjustbox}
\usepackage{multirow}
\usepackage{setspace}
\usepackage{hyperref}
\usepackage{comment}

\long\def\comment #1\commentend{}

\begin{document}

\title{\Large A Stochastic Epidemiological Model of Latent Tuberculosis in a Radiation Exposed Mars Colony}

\author{Teddy Lazebnik$^{1,2,*}$\\
\(^1\) Department of Information Systems, University of Haifa, Haifa, Israel \\
\(^2\) Department of Computing, Jonkoping University, Jonkoping, Sweden \\
\(^*\) Corresponding author: \url{teddy.lazebnik@ju.se} 
}

\date{ }

\maketitle 

\begin{abstract}
\noindent
Plans to establish a sustained human presence on Mars have moved from speculative ambition toward concrete engineering programmes, making the biological consequences of settlement an increasingly practical question. A Mars colony would place a small, closed population in an environment combining chronic radiation, altered immunity, constrained medical autonomy, and engineered indoor air. Latent infections are especially important because clinically silent carriers may become sources of transmissible disease when host control deteriorates. In this study, we develop a stochastic host–radiation–pathogen–habitat model of latent tuberculosis reactivation in a Mars colony. The model links galactic cosmic radiation to immune competence, immune competence to latent-tuberculosis reactivation, and reactivation to airborne transmission in a closed habitat. We also formulate countermeasure allocation as a partially observable sequential decision problem in which isolation and medication are selected by fixed baselines or by a proximal policy optimization policy trained on an agent-based simulator. Our simulations show that active tuberculosis can emerge endogenously despite no initial infectious cases, and that risk is most sensitive to latent reservoir size, radiation–immune coupling and reactivation sensitivity. Adaptive control reduced infectious burden and mortality while limiting unnecessary intervention. This framework supports mission-specific stress testing of screening, monitoring, shielding and treatment strategies before launch. \\ \\

\end{abstract}

\maketitle \thispagestyle{empty}
\pagestyle{myheadings} \markboth{Draft:  \today}{Draft:  \today}
\setcounter{page}{1}

\section{Introduction}
Human space exploration is transitioning from short-duration orbital missions toward exploration-class missions and eventual settlements beyond low Earth orbit, where radiation, altered gravity, confinement, isolation, and closed-loop life-support systems jointly shape human biological risk rather than acting as independent hazards \cite{afshinnekoo2020fundamental,hessel2022space,neukart2024towards}. Among the medical threats expected to become more consequential during long-duration missions, infections are distinctive because they can degrade individual health, spread through shared habitats, consume scarce medical resources, and compromise mission success \cite{cowen2024infections,cowen2024infections,castro2024infectious}. These concerns become sharper in the Mars-colony setting, where communication delays, limited evacuation options, restricted diagnostics, constrained pharmacy mass, and dependence on crew medical autonomy make even familiar terrestrial infections operationally unfamiliar \cite{tran2021evidence,zaccaria2025effects,khoshtinat2025earth}. Unlike terrestrial outbreaks, a Mars-colony epidemic would occur in a small, closed, highly structured population whose air, surfaces, schedules, waste streams, and medical resources are engineered parts of the transmission environment \cite{mora2016confined,lazebnik2023roomAirborne,alexi2022ventilationTradeoff}. In parallel, the colony would be exposed to a radiation environment dominated by galactic cosmic rays and episodic solar particle events, requiring risk assessment that links habitat shielding, mission timing, and biological response \cite{slaba2025validated,townsend2018storm}. Measurements from the Mars Science Laboratory Radiation Assessment Detector showed that the Martian surface has a persistent ionizing-radiation environment modulated by atmosphere, heliophysics, and secondary-particle production \cite{hassler2014mars,kennedy2014biological}. Measurements during the cruise phase to Mars further demonstrated that interplanetary transit itself can contribute a substantial fraction of total mission radiation exposure, making the outbound and return phases relevant to colony health even before permanent surface habitation begins \cite{zeitlin2013measurements}.

The uniqueness of pandemics in space is therefore not simply that pathogens may be present, but that host susceptibility, pathogen behavior, and the built environment are all perturbed at the same time \cite{winer2026astroimmunology}. The National Aeronautics and Space Administration's (NASA) microbiology technical guidance explicitly frames spacecraft microbial contamination as a risk to crew health and onboard systems, noting that inflight infectious disease occurs, that some spaceflight conditions can alter microbial virulence, and that astronaut immune responses are modified during flight \cite{nasa2024microbiology}. NASA's Human Research Roadmap similarly identifies immune dysregulation, altered microbial virulence, and host--microorganism interactions as research gaps for exploration missions \cite{nasa2026hrrimmune}. Operationally, NASA has historically mitigated acute preflight infection importation through health-stabilization procedures, and a retrospective analysis of missions from Apollo through 2024 provides evidence that such programs reduce, but do not conceptually eliminate, infection-related risk \cite{blue2026hsp}. At the quantitative-risk level, NASA's Integrated Medical Model uses probabilistic Monte Carlo methods to estimate medical events, resource needs, and mission impacts across spaceflight scenarios \cite{antonsen2022estimating}. More recent NASA work on the Medical Extensible Dynamic Probabilistic Risk Assessment Tool indicates a move toward higher-fidelity, event-based health-risk simulation for exploration missions \cite{prelich2024medprat}. However, probabilistic medical-event models do not by themselves resolve the epidemiological dynamics of an infection that begins silently in one colonist, becomes clinically active under altered host immunity, and then spreads through shared air in a closed settlement \cite{menzies2018progression}.

Pathogens can reach space through several routes, including crew microbiomes, cargo, food systems, water systems, environmental surfaces, and latent infections carried by apparently healthy individuals \cite{checinska2019isssurfaces}. Longitudinal astronaut microbiome studies show that spaceflight can alter microbial communities across gastrointestinal, skin, nasal, and oral body sites, linking the crew itself to the habitat microbiome \cite{voorhies2019impact}. Evidence from long-duration missions also shows that latent herpesviruses can reactivate and shed during spaceflight, making endogenous infection reservoirs a documented spaceflight phenomenon rather than a purely theoretical concern \cite{mehta2017latent}. A later synthesis of herpesvirus reactivation in astronauts reported frequent viral shedding during Shuttle and International Space Station missions and associated this pattern with stress-axis activation and reduced cell-mediated immunity \cite{rooney2019herpes}. Latent tuberculosis infection provides a different but epidemiologically sharper endogenous reservoir because \say{Mycobacterium tuberculosis} can remain clinically silent for years and later progress to active disease, with progression risk strongly shaped by immunological status \cite{who2018ltbi}. Public-health guidance distinguishes latent tuberculosis infection from active tuberculosis disease by noting that latent infection is not infectious, whereas reactivation can lead to transmissible disease \cite{cdc2024latent}. Active pulmonary or laryngeal tuberculosis is transmitted through airborne particles generated during coughing, speaking, or singing, making shared indoor air a central transmission pathway \cite{cdc2025spread}. Clinically, tuberculosis spans a spectrum from immunological containment within granulomas to symptomatic contagious pulmonary disease, making it well suited for models that couple host control, reactivation, and transmission \cite{pai2016tuberculosis}. Late reactivation is especially relevant to Mars settlement because systematic evidence indicates that tuberculosis can reactivate years after initial infection \cite{behr2021reactivation}. Current latent-tuberculosis diagnostics are useful for identifying infection but have limited ability to predict which infected individuals will progress to active disease, creating a residual screening-failure problem for long-duration settlements \cite{rangaka2012predictive}.

Recent space-biomedicine studies have substantially improved the empirical basis for modeling this problem, but they have mostly characterized components rather than closed-population epidemic mechanisms \cite{garrettbakelman2019twins}. The Inspiration4 longitudinal multi-omics study profiled host microbiome architecture and immune responses during short-duration spaceflight, showing that high-resolution microbial and immune time series are now feasible in civilian spaceflight contexts \cite{tierney2024longitudinal}. A companion single-cell multi-ome study identified conserved and sex-specific immune responses to spaceflight, including inflammatory markers, immune-gene signatures, and links between immune pathways and microbiome shifts \cite{kim2024singlecell}. Environmental studies of the International Space Station have also advanced from sparse monitoring toward spatially resolved habitat ecology, including a large 3D microbial and chemical map of the United States Orbital Segment \cite{salido2025iss}. Genomic and metabolic work on multidrug-resistant \emph{Enterobacter bugandensis} isolated from the ISS shows that clinically relevant organisms can persist, adapt, and participate in microbial succession within spacecraft environments \cite{singh2024bugandensis}. Analyses of bacteriophages in ISS bacterial genomes further suggest that microbial adaptation to spaceflight may involve functions related to antimicrobial resistance, virulence, DNA repair, and dormancy \cite{irby2024phage}. Standardized human-health monitoring frameworks such as Spaceflight Standard Measures are beginning to make astronaut physiological and omics datasets more comparable across missions \cite{hardy2025standardmeasures}. Nevertheless, recent medical-microbiology discussions of Mars-bound infection management still emphasize practical gaps in diagnostics, epidemiological data, antimicrobial decision-making, and mission-specific clinical breakpoints \cite{boschert2025spaceflight}.

Overall, there is an absence of a mechanistic epidemiological model that links Mars radiation exposure, immune suppression, latent tuberculosis reactivation, and airborne transmission inside a growing extraterrestrial colony. Existing tuberculosis models have examined progression from latent infection to active disease in terrestrial populations, but their assumptions generally do not include radiation-mediated immune perturbation, engineered habitat mixing, solar-particle-event sheltering, or multi-wave Mars settlement. To this end, in this study, we develop a stochastic host-radiation-pathogen-habitat model for latent tuberculosis in a Mars colony. 

The remainder of the paper is organized as follows. Section~2 formally defines the proposed host--radiation--pathogen--habitat model. Section~3 presents the optimal-control framework, describing the agent-based simulator and the partially observable countermeasure allocation model. Section~4 describes the experimental setup and reports the simulation results. Finally, Section~5 discusses the implications of the findings for Mars-colony health planning, summarizes the study limitations, and outlines future directions. 

\section{Model Definition}
The Mars-colony epidemiological model considers a closed colony with a fixed number of individuals \(N\). The colony is assumed to be well mixed, such that every living individual has the same average probability of epidemiological contact with every other living individual. This assumption follows the classical deterministic compartmental modeling framework introduced for epidemic dynamics and later generalized into SIR, SEIR, and SIRD-type models \cite{kermack1927contribution,lazebnik2023extendedSIRReview,hethcote2000mathematics}. Unlike a standard SIRD model, however, the proposed model includes a latent tuberculosis state and an immune-mediated reactivation process, consistent with tuberculosis models that distinguish latent infection from active infectious disease \cite{menzies2018progression,pai2016tuberculosis}.

\subsection{Epidemiological model}
Formally, each individual belongs to one of six epidemiological states: susceptible (\(S\)), latent tuberculosis infection (\(L\)), progressive active tuberculosis (\(P\)), infectious active tuberculosis (\(I\)), recovered or successfully treated (\(R\)), and dead (\(D\)). Susceptible individuals (\(S\)) have not been infected with \textit{Mycobacterium tuberculosis}. Latent individuals (\(L\)) carry tuberculosis infection but are not infectious. Progressive active individuals (\(P\)) represent individuals whose latent infection has reactivated, or whose new infection is progressing toward active disease, but who are not yet represented as infectious in the colony-level model. Infectious individuals (\(I\)) have active pulmonary tuberculosis and can transmit infection. Recovered individuals (\(R\)) are no longer infectious after successful treatment or immune control. Dead individuals (\(D\)) died from active tuberculosis or tuberculosis-associated complications. Therefore, at any time (\(t\)), \(N = S(t) + L(t) + P(t) + I(t) + R(t) + D(t)\). 

\begin{figure}[!ht]
    \centering
    \includegraphics[width=0.99\linewidth]{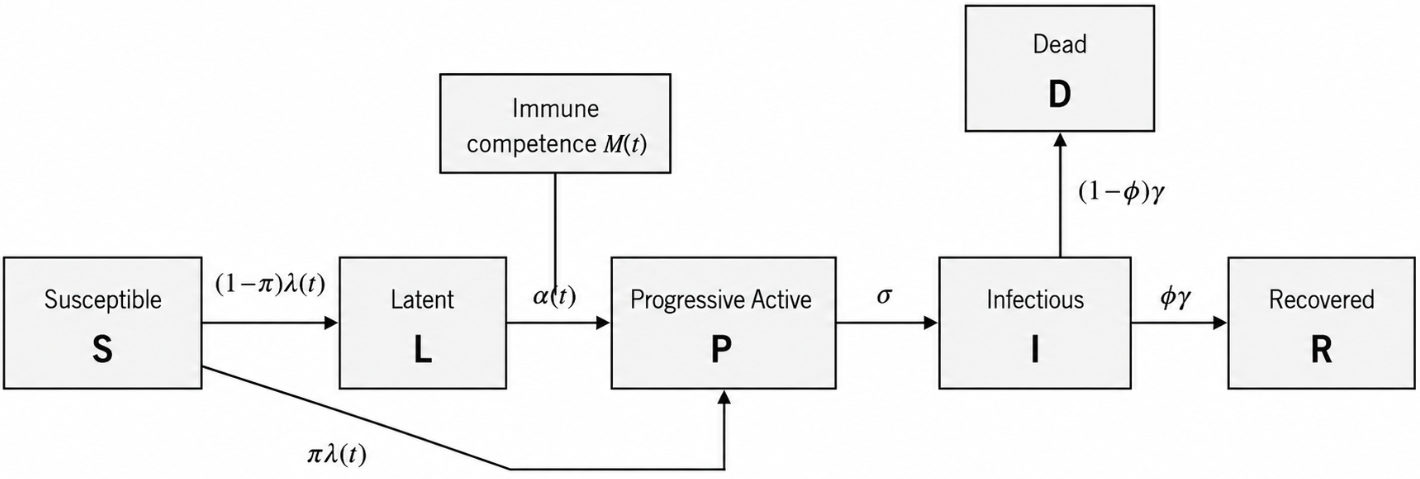}
    \caption{A schematic view of the transition between epidemiological states.}
    \label{fig:scheme}
\end{figure}

Susceptible individuals (\(S\)) become infected through contact with infectious individuals (\(I\)). A newly infected individual enters the latent state (\(L\)) with probability (\(1-\pi\)), or the progressive active state (\(P\)) with probability (\(\pi\)). Latent individuals (\(L\)) may reactivate into progressive active tuberculosis (\(P\)) at a time-dependent rate (\(\alpha(t)\)). Individuals in (\(P\)) become infectious at rate (\(\sigma\)). Infectious individuals (\(I\)) leave the infectious state at rate (\(\gamma\)). A fraction (\(\phi\)) of these individuals recover and move to (\(R\)), while the remaining fraction (\(1-\phi\)) die and move to (\(D\)). Fig. \ref{fig:scheme} presents a schematic view of the transition between epidemiological states. 

The force of infection is defined using standard frequency-dependent transmission \(\lambda(t) = \beta \frac{I(t)}{N-D(t)}\), where \(\beta\) is the effective transmission rate and \(N-D(t)\) is the number of living individuals in the colony. The denominator excludes dead individuals because they no longer participate in epidemiological contact. This form is a standard mass-action incidence assumption for well-mixed compartmental epidemic models \cite{brauer2019mathematical,lazebnik2023pandemicManagement}. We define \(M(t)\in[0,1]\) as the average immune competence of the colony. The value \(M(t)=1\) represents baseline pre-mission immune competence, while smaller values indicate immune suppression. Immune suppression is driven by chronic galactic cosmic radiation exposure. The immune state evolves according to
\begin{equation}
\frac{dM(t)}{dt}
=
\rho_M \big(1-M(t)\big)
-
\kappa_G G(t),
\label{eq:immune_ode}
\end{equation}
where \(\rho_M\) is the immune recovery rate, \(G(t)\) is chronic galactic cosmic radiation exposure, and \(\kappa_G\) determines how strongly chronic galactic cosmic radiation reduces immune competence. The first term, \(\rho_M(1-M(t))\), represents recovery of immune competence toward the baseline value \(M(t)=1\), while the second term, \(\kappa_G G(t)\), represents radiation-associated immune suppression. The use of immune suppression as a driver of latent-pathogen reactivation is motivated by spaceflight studies reporting immune dysregulation and latent viral reactivation in astronauts \cite{mehta2017latent,rooney2019herpes,sanzari2015leukocyte}.

The latent tuberculosis reactivation rate is modeled as a multiplicative hazard that increases as immune competence decreases,
\begin{equation}
\alpha(t)
=
\alpha_0
\exp\left(\theta \big(1-M(t)\big)\right),
\label{eq:reactivation_rate}
\end{equation}
where \(\alpha_0\) is the baseline latent tuberculosis reactivation rate under normal immune competence, and \(\theta\geq 0\) controls the sensitivity of reactivation to immune suppression. The exponential form follows the common proportional-hazards modeling idea that covariates multiply a baseline transition hazard \cite{cox1972regression}. In this model, radiation does not directly create tuberculosis infection; rather, chronic galactic cosmic radiation reduces immune competence, which increases the rate at which latent tuberculosis becomes active.

The epidemiological dynamics are described by Eqs.~\eqref{eq:S_ode}--\eqref{eq:D_ode}. In Eq.~\eqref{eq:S_ode}, \(\frac{dS(t)}{dt}\) is the change in the number of susceptible individuals over time. Susceptible individuals decrease only when they are infected by infectious individuals. The infection rate is determined by the force of infection \(\lambda(t)\).

\begin{equation}
\frac{dS(t)}{dt}
=
-\lambda(t)S(t).
\label{eq:S_ode}
\end{equation}

In Eq.~\eqref{eq:L_ode}, \(\frac{dL(t)}{dt}\) is the change in the number of individuals with latent tuberculosis infection. This group increases when susceptible individuals become infected and enter latency with probability \(1-\pi\). It decreases when latent individuals reactivate and move to the progressively active tuberculosis state at rate \(\alpha(t)\).

\begin{equation}
\frac{dL(t)}{dt}
=
(1-\pi)\lambda(t)S(t)
-
\alpha(t)L(t).
\label{eq:L_ode}
\end{equation}

In Eq.~\eqref{eq:P_ode}, \(\frac{dP(t)}{dt}\) is the change in the number of individuals with progressive active tuberculosis. This group increases through two pathways. First, newly infected susceptible individuals enter progressive active tuberculosis directly with probability \(\pi\). Second, latent individuals reactivate at rate \(\alpha(t)\). The group decreases when progressive active individuals become infectious at rate \(\sigma\).

\begin{equation}
\frac{dP(t)}{dt}
=
\pi\lambda(t)S(t)
+
\alpha(t)L(t)
-
\sigma P(t).
\label{eq:P_ode}
\end{equation}

In Eq.~\eqref{eq:I_ode}, \(\frac{dI(t)}{dt}\) is the change in the number of infectious active tuberculosis cases. This group increases when progressive active individuals become infectious at rate \(\sigma\). It decreases when infectious individuals leave the infectious state at rate \(\gamma\), either because they recover after treatment or immune control, or because they die.

\begin{equation}
\frac{dI(t)}{dt}
=
\sigma P(t)
-
\gamma I(t).
\label{eq:I_ode}
\end{equation}

In Eq.~\eqref{eq:R_ode}, \(\frac{dR(t)}{dt}\) is the change in the number of recovered or successfully treated individuals. A fraction \(\phi\) of infectious individuals leaving the infectious state recover, such that the recovery flow is \(\phi\gamma I(t)\).

\begin{equation}
\frac{dR(t)}{dt}
=
\phi\gamma I(t).
\label{eq:R_ode}
\end{equation}

In Eq.~\eqref{eq:D_ode}, \(\frac{dD(t)}{dt}\) is the change in the number of dead individuals. A fraction \(1-\phi\) of infectious individuals leaving the infectious state die, such that the mortality flow is \((1-\phi)\gamma I(t)\).

\begin{equation}
\frac{dD(t)}{dt}
=
(1-\phi)\gamma I(t).
\label{eq:D_ode}
\end{equation}

Combining Eqs.~\eqref{eq:S_ode}--\eqref{eq:D_ode}, the complete epidemiological model is

\begin{equation}
\begin{array}{c}
\frac{dS(t)}{dt}
=
-\lambda(t)S(t),\;\;
\frac{dL(t)}{dt}
=
(1-\pi)\lambda(t)S(t)
-
\alpha(t)L(t),\;\;
\frac{dP(t)}{dt}
=
\pi\lambda(t)S(t)
+
\alpha(t)L(t)
-
\sigma P(t),\\\\
\frac{dI(t)}{dt}
=
\sigma P(t)
-
\gamma I(t),\;\;
\frac{dR(t)}{dt}
=
\phi\gamma I(t),\;\;
\frac{dD(t)}{dt}
=
(1-\phi)\gamma I(t).
\end{array}
\label{eq:full_epidemic_model}
\end{equation}

The general initial conditions are
\[
S(0)=S_0,\;\;
L(0)=L_0,\;\;
P(0)=P_0,\;\;
I(0)=I_0,\;\;
R(0)=R_0,\;\;
D(0)=0,\;\;
M(0)=M_0.
\]
However, for a Mars-colony scenario focused on endogenous reactivation, one may set \(I_0=P_0=0\) and \(L_0>0\). This represents a colony that begins with no active tuberculosis cases but contains one or more individuals with latent tuberculosis infection. In such a case, the first active case can only arise through the reactivation term \(\alpha(t)L(t)\), making immune suppression the initiating mechanism for the outbreak. 

\subsection{Countermeasures}
We consider two post-reactivation countermeasures: isolation and medication. These interventions are activated after active tuberculosis appears in the colony and therefore act on the transmission and clinical-progression terms rather than on the initial latent reservoir \cite{cdc2023tbinfectioncontrol,who2025tbtreatmentcare}. Formally, let \(u_I(t)\in[0,1]\) denote the intensity of isolation at time \(t\), where \(u_I(t)=0\) indicates no isolation and \(u_I(t)=1\) indicates maximal isolation. Isolation reduces the effective transmission rate according to \(\beta_I(t) = \beta\big(1-\eta_I u_I(t)\big)\), where \(\eta_I\in[0,1]\) is the maximal effectiveness of isolation. If \(\eta_I=0\), isolation has no effect on transmission. If \(\eta_I=1\), maximal isolation can fully eliminate transmission from infectious individuals in the model. The force of infection under isolation is therefore

\begin{equation}
\lambda_I(t)
=
\beta_I(t)\frac{I(t)}{N-D(t)}
=
\beta\big(1-\eta_I u_I(t)\big)\frac{I(t)}{N-D(t)}.
\label{eq:isolation_lambda}
\end{equation}

Medication is represented by a second control variable \(u_M(t)\in[0,1]\), where \(u_M(t)=0\) indicates no active medication and \(u_M(t)=1\) indicates maximal medication intensity. Medication affects the model through two mechanisms. First, it increases the rate at which infectious individuals leave the infectious state \(\gamma_M(t) = \gamma\big(1+\eta_\gamma u_M(t)\big)\), where \(\eta_\gamma\geq 0\) is the maximal proportional increase in the removal rate due to medication. Second, medication increases the probability that an infectious individual recovers rather than dies \(\phi_M(t)
=
\phi
+
\eta_\phi u_M(t)(1-\phi)\), where \(\eta_\phi\in[0,1]\) is the maximal improvement in survival due to medication. Under this formulation, \(\phi_M(t)=\phi\) when no medication is used, and \(\phi_M(t)\leq 1\) for all \(u_M(t)\in[0,1]\).

Using isolation and medication, the controlled epidemiological dynamics are

\begin{equation}
\begin{array}{c}
\frac{dS(t)}{dt}
=
-\lambda_I(t)S(t),\;\;

\frac{dL(t)}{dt}
=
(1-\pi)\lambda_I(t)S(t)
-
\alpha(t)L(t),\;\;

\frac{dP(t)}{dt}
=
\pi\lambda_I(t)S(t)
+
\alpha(t)L(t)
-
\sigma P(t),\\\\

\frac{dI(t)}{dt}
=
\sigma P(t)
-
\gamma_M(t) I(t),\;\;

\frac{dR(t)}{dt}
=
\phi_M(t)\gamma_M(t) I(t),\;\;

\frac{dD(t)}{dt}
=
\big(1-\phi_M(t)\big)\gamma_M(t) I(t).

\end{array}
\label{eq:controlled_model}
\end{equation}

The use of both countermeasures is assumed to be costly. Isolation can reduce crew availability, impose psychological burden, and consume protected living space, while medication consumes limited medical stock, requires diagnostic confidence, and may produce side effects or drug-management constraints. Therefore, a countermeasure burden term takes the form \cite{lazebnik2022npiEconomy,lazebnik2023borderClosure}:
\begin{equation}
\mathcal{B}(t)
=
c_I u_I(t)^2 I(t)
+
c_M u_M(t)^2 I(t),
\label{eq:countermeasure_burden}
\end{equation}
where \(c_I\) and \(c_M\) are the burden weights associated with isolation and medication, respectively. The quadratic form penalizes high-intensity countermeasure use more strongly than low-intensity use, reflecting the idea that strict isolation or aggressive medication becomes disproportionately costly in a constrained Mars-colony environment. Multiplication by \(I(t)\) ensures that the burden is incurred primarily when countermeasures are applied to infectious individuals.

If medication stock is limited, an additional resource constraint can be imposed:
\begin{equation}
\int_0^T u_M(t)I(t),dt
\leq
Q_M,
\label{eq:medication_constraint}
\end{equation}
where \(Q_M\) is the total available medication capacity over the planning horizon \(T\). Similarly, if isolation space or crew-support capacity is limited, one may impose
\begin{equation}
\int_0^T u_I(t)I(t),dt
\leq
Q_I,
\label{eq:isolation_constraint}
\end{equation}
where \(Q_I\) is the total available isolation capacity. These constraints force the colony to use countermeasures strategically rather than applying maximal isolation and medication at all times.

The policy objective is to minimize disease burden, mortality, and countermeasure burden simultaneously:
\begin{equation}
\min_{u_I(t),u_M(t)}
\left[
w_I\int_0^T I(t),dt
+
w_D D(T)
+
w_B\int_0^T \mathcal{B}(t),dt
\right],
\label{eq:countermeasure_objective}
\end{equation}
subject to \(0\leq u_I(t)\leq 1\), \(0\leq u_M(t)\leq 1\), and the resource constraints in Eqs.~\eqref{eq:medication_constraint}--\eqref{eq:isolation_constraint}. Here, \(w_I\), \(w_D\), and \(w_B\) are non-negative weights assigned to infectious disease burden, mortality, and countermeasure burden, respectively. This formulation captures the operational tradeoff faced by a Mars colony where strong isolation and medication can suppress an outbreak, but excessive use may consume scarce resources and reduce mission functionality \cite{lazebnik2022npiEconomy,lazebnik2023hospitalDRL}.

\subsection{Outcome metrics}
In order to evaluate the epidemiological and operational consequences of latent tuberculosis reactivation in a Mars colony, we measure three model outcomes. The first captures how often latent or newly acquired infection progresses into active disease. The second captures how much infectious disease burden the colony experiences. The third captures the operational cost of using isolation and medication \cite{keeling2008modeling,sharomi2017optimal}.

The first outcome is the cumulative number of individuals who enter progressive active tuberculosis during the time horizon \(T\). This metric is important because the central risk in the proposed Mars-colony scenario is not only direct transmission, but the activation of a silent latent reservoir. We define the cumulative active tuberculosis incidence \(C_A(t)\) by
\begin{equation}
\frac{dC_A(t)}{dt}
=
\pi\lambda_I(t)S(t)
+
\alpha(t)L(t),
\qquad
C_A(0)=0.
\label{eq:cumulative_active_incidence}
\end{equation}
where the first term, \(\pi\lambda_I(t)S(t)\), represents newly infected susceptible individuals who progress directly into the progressive active state \(P\). The second term, \(\alpha(t)L(t)\), represents latent tuberculosis reactivation. 

The second outcome measures the clinical and transmission burden caused by active infectious tuberculosis. We define the infectious person-time burden as
\begin{equation}
B_I(T)
=
\int_0^T I(t),dt.
\label{eq:infectious_burden}
\end{equation}
This metric represents the total amount of time the colony spends with infectious active tuberculosis present. It is relevant because transmission opportunity, clinical workload, monitoring demand, and operational disruption all increase with both the number of infectious individuals and the duration of infectiousness. We also measure the peak infectious burden,
\begin{equation}
I_{\max}
=
\max_{0\leq t\leq T} I(t),
\label{eq:peak_infectious_burden}
\end{equation}
which captures the maximum simultaneous infectious load that the colony must manage. This is especially important in a small Mars colony, where even one or two active infectious cases may represent a substantial medical and operational burden.

The third outcome measures the burden of using isolation and medication. This metric is necessary because, in a resource-limited Mars colony, maximal use of all countermeasures at all times is unlikely to be operationally feasible. Isolation may reduce available crew labor and consume protected living space, while medication may consume limited pharmaceutical stock and require medical supervision. Therefore, we define the cumulative countermeasure burden as

\begin{equation}
B_U(T)
=
\int_0^T
\left[
c_Iu_I(t)^2I(t)
+
c_Mu_M(t)^2I(t)
\right]dt,
\label{eq:countermeasure_burden_metric}
\end{equation}
where \(u_I(t)\in[0,1]\) is the isolation intensity, \(u_M(t)\in[0,1]\) is the medication intensity, and \(c_I,c_M\geq 0\) are the respective burden weights. The quadratic form penalizes high-intensity intervention more strongly than low-intensity intervention, which is common in optimal-control formulations of biological and epidemiological systems \cite{lenhart2007optimal}. 

For interpretation, we also report the controlled colony reproduction indicator:
\begin{equation}
\mathcal{R}*c(t)
=
\frac{\beta\big(1-\eta_Iu_I(t)\big)}
{\gamma\big(1+\eta*\gamma u_M(t)\big)}
\frac{S(t)}{N-D(t)}.
\label{eq:controlled_reproduction_indicator}
\end{equation}
This quantity is not an additional objective term, but it helps interpret whether the current combination of isolation and medication is expected to suppress transmission. Values \(\mathcal{R}_c(t)<1\) indicate that active transmission is expected to decline under the current conditions, whereas \(\mathcal{R}_c(t)>1\) indicates that an infectious case may generate more than one secondary infection in the well-mixed colony. Reproduction-number thresholds are a standard tool for interpreting epidemic growth and control in compartmental models \cite{diekmann2010construction}.

\section{Optimal control}
The deterministic model in the previous section defines the average epidemiological dynamics of latent tuberculosis reactivation and transmission in a well-mixed Mars colony. However, for the purpose of countermeasure allocation, a deterministic Ordinary Differential Equation (ODE) model has two limitations. First, the colony population may be small enough that stochastic effects, especially the timing of the first reactivation event, can strongly influence the outbreak trajectory. Second, isolation and medication are not merely continuous biological rates; they are operational decisions that must be allocated over time under resource and burden constraints. Therefore, we formulate the countermeasure problem as a sequential decision problem in which an agent observes the colony state and chooses isolation and medication intensities. The epidemiological dynamics are generated by an agent-based simulation (ABS), and the control policy is learned using deep reinforcement learning \cite{lazebnik2023hospitalDRL,shuchami2025warPandemicDRL}.

\subsection{Agent-based simulation implementation}
We construct an ABS that is consistent with the compartmental model but represents each colonist as an individual stochastic unit \cite{bonabeau2002agentbased}. Let \(\mathcal{A}={1,\dots,N}\) denote the set of colonists. Each colonist \(a\in\mathcal{A}\) is assigned an epidemiological state \(X_a(t)\in{S,L,P,I,R,D}\), where the states have the same meaning as in the compartmental model. The aggregate compartment sizes are obtained by summing over agents:
\[\forall \mathbb{S}\in\{S, L, P,I, R, D\}:  \mathbb{S}(t)=\sum_{a=1}^{N}\mathbb{I}[X_a(t)=\mathbb{S}].
\]

At each simulation step of length \(\Delta t\), living susceptible agents are exposed to the force of infection
\(
\lambda_I(t) = \beta\big(1-\eta_Iu_I(t)\big) \frac{I(t)}{N-D(t)},
\)
where \(u_I(t)\in[0,1]\) is the isolation intensity chosen by the controller. A susceptible agent becomes infected during the interval \([t,t+\Delta t)\) with probability
\(
p_{S\rightarrow {L,P}}(t)
=
1-\exp\big(-\lambda_I(t)\Delta t\big).
\)

If infection occurs, the agent enters the latent state \(L\) with probability \(1-\pi\) and the progressive active state (P) with probability \(\pi\). Thus, the stochastic infection transition is
\[
S
\rightarrow
\begin{cases}
L, & \text{with probability } (1-\pi)p_{S\rightarrow {L,P}}(t),\\
P, & \text{with probability } \pi p_{S\rightarrow {L,P}}(t).
\end{cases}
\]

Latent agents reactivate according to the time-dependent reactivation rate \(\alpha(t)\). The probability that a latent agent enters the progressive active state during one simulation step is
\(
p_{L\rightarrow P}(t)
=
1-\exp\big(-\alpha(t)\Delta t\big),
\)
where
\(
\alpha(t)
=
\alpha_0
\exp\left(\theta(1-M(t))\right).
\)

The immune-competence variable \(M(t)\) is updated at the colony level according to
\(
M(t+\Delta t)
=
M(t)
+
\Delta t
\left[
\rho_M(1-M(t))
-
\kappa_GG(t)
\right].
\)

After the update, \(M(t+\Delta t)\) is truncated to the interval \([0,1]\). This ensures that immune competence remains biologically interpretable. Progressive active agents become infectious with probability
\(
p_{P\rightarrow I}
=
1-\exp(-\sigma\Delta t).
\)
Infectious agents leave the infectious state according to the medication-controlled removal rate
\(
\gamma_M(t)
=
\gamma\big(1+\eta_\gamma u_M(t)\big),
\)
where \(u_M(t)\in[0,1]\) is the medication intensity chosen by the controller. The probability that an infectious agent leaves the infectious state during one simulation step is
\(
p_{I\rightarrow {R,D}}(t)
=
1-\exp\big(-\gamma_M(t)\Delta t\big).
\)
Conditional on leaving the infectious state, the agent recovers with probability
\(
\phi_M(t)
=
\phi+\eta_\phi u_M(t)(1-\phi),
\)
and dies with probability \(1-\phi_M(t)\). Therefore,
\[
I
\rightarrow
\begin{cases}
R, & \text{with probability } \phi_M(t)p_{I\rightarrow {R,D}}(t),\\
D, & \text{with probability } \big(1-\phi_M(t)\big)p_{I\rightarrow {R,D}}(t).
\end{cases}
\]
The simulation is initialized with \(I(0)=P(0)=0\) and \(L(0)>0\) in the endogenous reactivation scenario. 

\subsection{Deep reinforcement learning for countermeasure allocation under partial observability}
We formulate countermeasure allocation as a sequential decision-making problem in which the controller chooses isolation and medication intensities over time. However, unlike the ABS, the controller does not observe the true epidemiological state of every colonist as it is sampled from the colonists' suits' biometric indicators. This results in a partially observable control problem, which can be interpreted as a partially observable Markov decision process (MDP) \cite{kaelbling1998planning,sutton2018reinforcement}.

For each colonist \(a\in{1,\dots,N}\), the true epidemiological state is \(X_a(t)\in{S,L,P,I,R,D}\). The monitoring system does not distinguish susceptible, latent, and progressive active individuals. Hence, all individuals in \(S\), \(L\), and \(P\) appear identical to the controller and are observed as \(\widetilde{S}\): \(X_a(t)\in{S,L,P}
\quad\Longrightarrow\quad
Y_a(t)=\widetilde{S}\). Thus, \(\widetilde{S}\) should not be interpreted as true susceptibility. Rather, it represents all individuals who are not detected as infectious, recovered, or dead. In addition, the controller does not observe the immune competence variable \(M(t)\). This means that the controller cannot directly observe the biological mechanism that increases the latent reactivation rate \(\alpha(t)\).

The infectious state is detected probabilistically through the colony monitoring system. Let \(\tau_a^I(t)\) denote the number of consecutive time steps for which colonist \(a\) has been in the true infectious state \(I\). We define \(\zeta_I\in(0,1)\) as the one-step probability that a truly infectious individual remains undetected. Therefore, the probability that an infectious colonist remains undetected after \(\tau_a^I(t)\) time steps is \(
\Pr\left(Y_a(t)\neq\widetilde{I}\mid X_a(t)=I\right)
=
\zeta_I^{\tau_a^I(t)}\). Equivalently, the probability that an infectious colonist is detected by time \(t\) is \(\Pr\left(Y_a(t)=\widetilde{I}\mid X_a(t)=I\right)
=
1-\zeta_I^{\tau_a^I(t)}\). The recovered state is detected in the same manner. Let \(\tau_a^R(t)\) denote the number of consecutive time steps for which colonist \(a\) has been in the true recovered state \(R\), and let \(\zeta_R\in(0,1)\) denote the one-step probability that a truly recovered individual remains undetected. Then \(\Pr\left(Y_a(t)\neq\widetilde{R}\mid X_a(t)=R\right)
=
\zeta_R^{\tau_a^R(t)}\) and \(\Pr\left(Y_a(t)=\widetilde{R}\mid X_a(t)=R\right)
=
1-\zeta_R^{\tau_a^R(t)}\).
Dead individuals are assumed to be observed without uncertainty:
\(X_a(t)=D\quad\Longrightarrow\quad Y_a(t)=D\).

The PPO policy receives the observation vector
\begin{equation}
\mathbf{o}_t
=
\left[
\frac{\widetilde{S}(t)}{N},
\frac{\widetilde{I}(t)}{N},
\frac{\widetilde{R}(t)}{N},
\frac{D(t)}{N},
q_I(t),
q_M(t)
\right],
\label{eq:ppo_partial_observation}
\end{equation}
where \(q_I(t)\) and \(q_M(t)\) are the remaining isolation and medication capacities, respectively. If explicit capacity depletion is not modeled, \(q_I(t)\) and \(q_M(t)\) are omitted. Importantly, the observation vector does not include \(L(t)\), \(P(t)\), true \(I(t)\), true \(R(t)\), or \(M(t)\). 

The action vector is \(\mathbf{a}_t
=
\left[
u_I(t),
u_M(t)
\right]\) such that \(u_I(t),u_M(t)\in[0,1]\), where \(u_I(t)\) is the isolation intensity and \(u_M(t)\) is the medication intensity. However, these countermeasures can only be applied to individuals detected as infectious. Thus, isolation and medication affect agents with \(Y_a(t)=\widetilde{I}\), not all agents whose true state is \(I\). In the individual-based simulator, this is implemented directly: if \(X_a(t)=I\) and \(Y_a(t)=\widetilde{I}\), then isolation and medication modify that agent's transition and transmission parameters; if \(X_a(t)=I\) but \(Y_a(t)=\widetilde{S}\), the individual remains undetected, non-isolated, and untreated during that time step.

The reward is computed from the true epidemiological outcomes because deaths and infectious burden represent the actual mission cost. However, the policy chooses actions using only the partially observed vector \(\mathbf{o}_t\). The immediate reward is
\begin{equation}
r_t
=
-
\left[
w_I\frac{I(t)}{N}
+
w_D\frac{\Delta D(t)}{N}
+
w_U
\left(
c_Iu_I(t)^2\frac{\widetilde{I}(t)}{N}
+
c_Mu_M(t)^2\frac{\widetilde{I}(t)}{N}
\right)
\right],
\label{eq:partial_reward_drl}
\end{equation}
where \(\Delta D(t)=D(t+\Delta t)-D(t)\). The first term penalizes the true infectious burden, the second term penalizes deaths, and the third term penalizes the operational burden of applying isolation and medication to detected infectious individuals. This formulation creates a realistic control problem: the controller is penalized for hidden infectious burden but can only act on detected cases.

The reinforcement-learning objective is to maximize the expected discounted return
\begin{equation}
J(\theta_{\pi})
=
\mathbb{E}*{\pi*{\theta_{\pi}}}
\left[
\sum_{t=0}^{T}
\delta_{RL}^{t} r_t
\right],
\label{eq:partial_rl_objective}
\end{equation}
where \(\pi_{\theta_{\pi}}(\mathbf{a}*t|\mathbf{o}*t)\) is the stochastic policy parameterized by neural-network weights \(\theta*{\pi}\), and \(\delta*{RL}\in(0,1]\) is the reinforcement-learning discount factor. We use \(\delta_{RL}\) rather than \(\zeta\) to avoid confusion with the monitoring parameters \(\zeta_I\) and \(\zeta_R\).

The policy is trained using Proximal Policy Optimization (PPO), a policy-gradient method that improves stability by clipping large policy updates \cite{schulman2017ppo}. Let
\[
r_t(\theta_{\pi})
=
\frac{
\pi_{\theta_{\pi}}(\mathbf{a}*t|\mathbf{o}*t)
}{
\pi*{\theta*{\pi}^{old}}(\mathbf{a}_t|\mathbf{o}_t)
}
\]
be the probability ratio between the updated policy and the previous policy. PPO maximizes the clipped surrogate objective
\begin{equation}
\mathcal{L}^{PPO}(\theta_{\pi})
=
\mathbb{E}*t
\left[
\min
\left(
r_t(\theta*{\pi})\hat{A}*t,
\mathrm{clip}
\left(
r_t(\theta*{\pi}),
1-\epsilon,
1+\epsilon
\right)
\hat{A}_t
\right)
\right],
\label{eq:ppo_partial_objective}
\end{equation}
where \(\hat{A}_t\) is an advantage-function estimator and \(\epsilon\) is the clipping parameter. In the present setting, PPO learns a mapping from imperfect monitoring information to countermeasure intensities. For example, the policy may learn to apply medication aggressively once \(\widetilde{I}(t)>0\), conserve medication when no infectious cases are detected, or combine isolation and medication when the detected infectious count persists across multiple time steps \cite{lazebnik2023hospitalDRL,alexi2023securityGames}.

Training proceeds by repeatedly simulating Mars-colony outbreak episodes. At each time step, the monitoring process maps the true agent states \(X_a(t)\) to observed states \(Y_a(t)\), the PPO policy observes \(\mathbf{o}_t\), selects \(\mathbf{a}_t=[u_I(t),u_M(t)]\), and the ABS advances the true epidemic state. 

\section{Experiments}
In this section, we outline \textit{in silico} experiments that are designed to evaluate both the epidemiological behavior of the proposed Mars-colony tuberculosis model and the effectiveness of adaptive countermeasure allocation. We first define the experimental setup, followed by the obtained results. 

\subsection{Setup}
Unless otherwise stated, time is measured in days, and the simulation horizon is \(T=1095\), corresponding to three Earth years. The default colony size is \(N=100\). The baseline scenario is an endogenous-reactivation scenario with \(L(0)>0\), \(P(0)=0\), and \(I(0)=0\). Thus, the colony initially contains latent tuberculosis carriers, but no active tuberculosis cases, and the first active case can arise only through the reactivation term \(\alpha(t)L(t)\).

In addition, the model assumes that chronic galactic cosmic radiation affects tuberculosis risk indirectly through immune competence rather than directly creating infection. We define (G(t)) as the effective chronic galactic cosmic radiation exposure rate after colony shielding. In the baseline experiments, we use a constant exposure rate \(G(t)=G_0(1-\eta_G)\), where \(G_0\) is the unshielded Mars surface radiation exposure rate and \(\eta_G\in[0,1]\) is the effective shielding level. In the baseline setup, \(\eta_G=0\), so \(G(t)=G_0\). The radiation immune-suppression function is defined as \(\Psi_G(t)=\kappa_G G(t)\), where \(\kappa_G\) converts chronic radiation exposure into loss of average immune competence. 

Table~\ref{tab:combined_parameters} summarizes all parameters used in the experiments. Parameters are divided into four groups: ODE epidemiological parameters, radiation and immune parameters, ABS parameters, and PPO control parameters. When a value is not directly available from the literature, it is marked as assumed and evaluated through sensitivity analysis.

\small
\begin{longtable}{p{0.15\textwidth} p{0.42\textwidth} p{0.20\textwidth} p{0.16\textwidth}}
\caption{The model's parameters with their default values.}
\label{tab:combined_parameters}\\

\hline \hline
\textbf{Parameter} & \textbf{Description} & \textbf{Default value} & \textbf{Source} \\
\hline \hline
\endfirsthead

\multicolumn{4}{l}{\textit{Table~\ref{tab:combined_parameters} continued from previous page.}}\\
\hline
\textbf{Parameter} & \textbf{Description} & \textbf{Default value} & \textbf{Source} \\
\hline
\endhead

\hline
\multicolumn{4}{r}{\textit{Continued on next page.}}\\
\endfoot

\hline
\endlastfoot

\multicolumn{4}{l}{\textbf{ODE epidemiological parameters}}\\
\hline
\(N\) & Initial colony size & \(100\) & Assumed \\
\(T\) & Simulation horizon & \(1095\) days & Assumed \\
\(\beta\) & Effective transmission rate & \(0.0083\) day\(^{-1}\) & \cite{ma2018quantifying} \\
\(\pi\) & Probability that a new infection enters \(P\) rather than \(L\) & \(0.05\) & \cite{menzies2018progression} \\
\(\alpha_0\) & Baseline LTBI reactivation rate under normal immune competence & \(2.8\times10^{-6}\) day\(^{-1}\) & \cite{cdc2024latent,who2018ltbi} \\
\(\theta\) & Sensitivity of reactivation to immune suppression & \(3\) & Assumed \\
\(\sigma\) & Rate from progressive active tuberculosis to infectious tuberculosis & \(1/90\) day\(^{-1}\) & \cite{menzies2018progression,pai2016tuberculosis} \\
\(\gamma\) & Baseline rate at which infectious individuals leave \(I\) & \(1/180\) day\(^{-1}\) & \cite{cdc2025dstreatment} \\
\(\phi\) & Probability of recovery after leaving \(I\) & \(0.88\) & \cite{who2024tbreport} \\

\hline
\multicolumn{4}{l}{\textbf{Radiation and immune parameters}}\\
\hline
\(G_0\) & Unshielded chronic Mars surface radiation exposure rate & \(0.64\) mSv day\(^{-1}\) & \cite{hassler2014mars} \\
\(\eta_G\) & Effective shielding level for chronic radiation & \(0\) & Assumed \\
\(\rho_M\) & Immune recovery rate toward baseline & \(1/180\) day\(^{-1}\) & Assumed \\
\(\kappa_G\) & Immune suppression per unit chronic radiation exposure & \(2.0\times10^{-3}\) mSv\(^{-1}\) & Assumed \\
\hline
\multicolumn{4}{l}{\textbf{Agent-based simulation parameters}}\\
\hline
\(\Delta t\) & Simulation time step & \(1\) day & Assumed \\
\(n_{eval}\) & Monte Carlo replications per evaluated policy & \(250\) & Assumed \\
-- & Transition probability from rate \(r(t)\) over one step & \(1-\exp[-r(t)\Delta t]\) & \cite{allen2017primer} \\
\hline
\multicolumn{4}{l}{\textbf{Monitoring and partial observability parameters}}\\
\hline
\(\zeta_I\) & One-step probability that a truly infectious individual remains undetected & \(0.85\) & Assumed \\
\(\zeta_R\) & One-step probability that a truly recovered individual remains undetected & \(0.85\) & Assumed \\

\hline
\multicolumn{4}{l}{\textbf{Countermeasure and PPO parameters}}\\
\hline
\(\eta_I\) & Maximum proportional reduction in transmission due to isolation & \(0.75\) & Assumed \\
\(\eta_\gamma\) & Maximum proportional increase in removal rate due to medication & \(1.0\) & Assumed \\
\(\eta_\phi\) & Maximum improvement in recovery probability due to medication & \(0.8\) & Assumed \\
\(c_I\) & Isolation burden weight & \(0.10\) & Assumed \\
\(c_M\) & Medication burden weight & \(0.10\) & Assumed \\
\(w_I\) & Infectious-burden reward weight & \(1\) & Assumed \\
\(w_D\) & Mortality reward weight & \(50\) & Assumed \\
\(w_U\) & Countermeasure-burden reward weight & \(1\) & Assumed \\
\(\delta_{RL}\) & RL discount factor & \(0.99\) & \cite{sutton2018reinforcement} \\
-- & Actor--critic hidden layers & \(2\) layers, \(64\) units each & \cite{raffin2021stablebaselines3} \\
-- & Learning rate & \(3\times10^{-4}\) & \cite{raffin2021stablebaselines3} \\
-- & Generalized Advantage Estimation (GAE) parameter & \(0.95\) & \cite{raffin2021stablebaselines3} \\
-- & Clip range & \(0.20\) & \cite{schulman2017ppo} \\
-- & Rollout length & \(2048\) steps & \cite{raffin2021stablebaselines3} \\
-- & Batch size & \(64\) & \cite{raffin2021stablebaselines3} \\
-- & Optimization epochs per rollout & \(10\) & \cite{raffin2021stablebaselines3} \\ \hline\hline

\end{longtable}
\normalsize

The experiments are organized into three complementary parts. First, we study the uncontrolled disease and immune-radiation dynamics in order to characterize when latent tuberculosis reactivation becomes epidemiologically important in a Mars-colony setting. In this part, no post-reactivation intervention is applied, i.e., \(u_I(t)=u_M(t)=0\). We use the baseline parameter set in Table~\ref{tab:combined_parameters} and then perform sensitivity analyses over the main uncertain and biologically relevant parameters. These include the initial number of latent tuberculosis carriers \(L(0)\), the transmission rate \(\beta\), the radiation exposure level \(G(t)\), the shielding level \(\eta_G\), the immune-recovery rate \(\rho_M\), the radiation immune-suppression coefficient \(\kappa_G\), and the reactivation sensitivity parameter \(\theta\). The goal of this experiment is to learn which parts of the coupled radiation--immune--tuberculosis system most strongly determine outbreak initiation, cumulative active tuberculosis incidence, peak infectious burden, mortality, and the timing of the first active case.

Second, we compare countermeasure strategies after active tuberculosis appears in the colony. We evaluate five policies under identical stochastic simulation conditions and identical initial states: no intervention, three baseline intervention policies, and the PPO-learned intervention policy. The no-intervention policy sets \(u_I(t)=u_M(t)=0\) throughout the simulation. The three baseline policies are: an isolation-only policy, in which detected infectious individuals are isolated with a fixed intensity; a medication-only policy, in which detected infectious individuals receive medication with a fixed intensity; and a combined fixed intervention policy, in which both isolation and medication are applied once at least one infectious individual is detected. These baseline policies provide interpretable reference strategies against which the PPO policy can be compared. The PPO policy chooses \(u_I(t)\) and \(u_M(t)\) adaptively from the partially observed colony state in Eq.~\eqref{eq:ppo_partial_observation}. All policies are evaluated using the same Monte Carlo seeds and are compared according to cumulative active tuberculosis incidence \(C_A(T)\), infectious person-time burden \(B_I(T)\), peak infectious burden \(I_{\max}\), deaths \(D(T)\), cumulative countermeasure burden \(B_U(T)\), and the policy objective in Eq.~\eqref{eq:countermeasure_objective}.

\begin{figure}[!ht]
    \centering
    \includegraphics[width=0.99\linewidth]{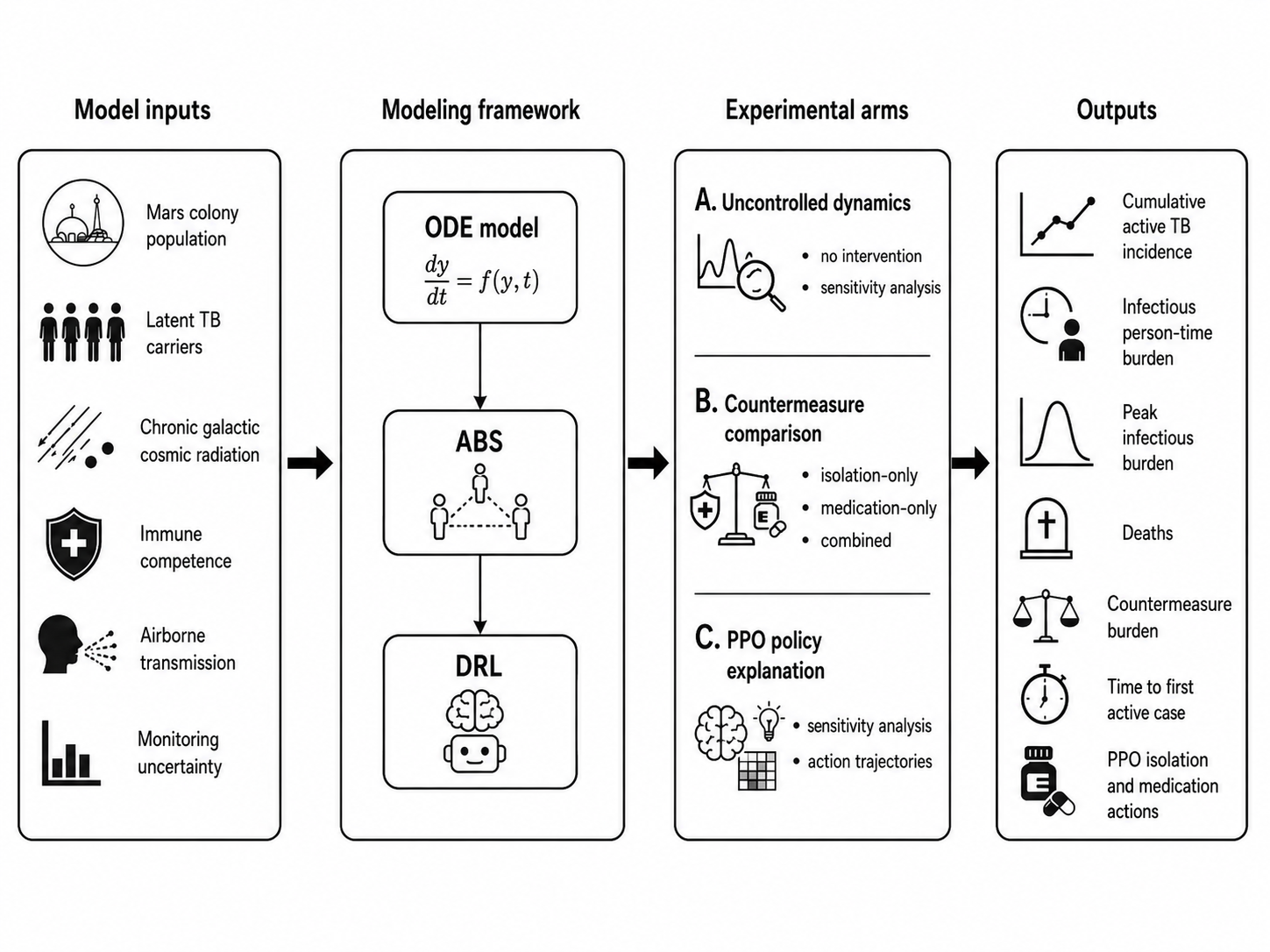}
    \caption{A schematic view of the experimental design.}
    \label{fig:process}
\end{figure}

Third, we examine how the learned intervention changes when the underlying dynamics change. This experiment serves both as a sensitivity analysis of the control policy and as an explanation of the PPO-learned intervention. We evaluate the trained PPO policy across parameter regimes that modify the epidemic, immune, and observation dynamics, including changes in \(\beta\), \(L(0)\), \(G(t)\), \(\eta_G\), \(\kappa_G\), \(\theta\), \(\rho_M\), \(\zeta_I\), and \(\zeta_R\). For each regime, we record not only epidemiological outcomes but also the action trajectories \(u_I(t)\) and \(u_M(t)\). We then summarize how the learned policy allocates isolation and medication as a function of detected infectious burden, remaining resources, and outbreak stage. This analysis allows us to determine whether the PPO policy behaves consistently with epidemiological intuition, for example by increasing isolation when the detected infectious burden rises, increasing medication when mortality risk is high, or conserving resources when detected cases are absent. In addition, we use action-response plots and policy heatmaps over the observed state variables to explain how the learned intervention adapts to changes in the hidden radiation-mediated reactivation dynamics. Fig. \ref{fig:process} provides a schematic view of the experimental design.

\subsection{Results}
\begin{figure}[!ht]
    \centering
    \includegraphics[width=0.99\linewidth]{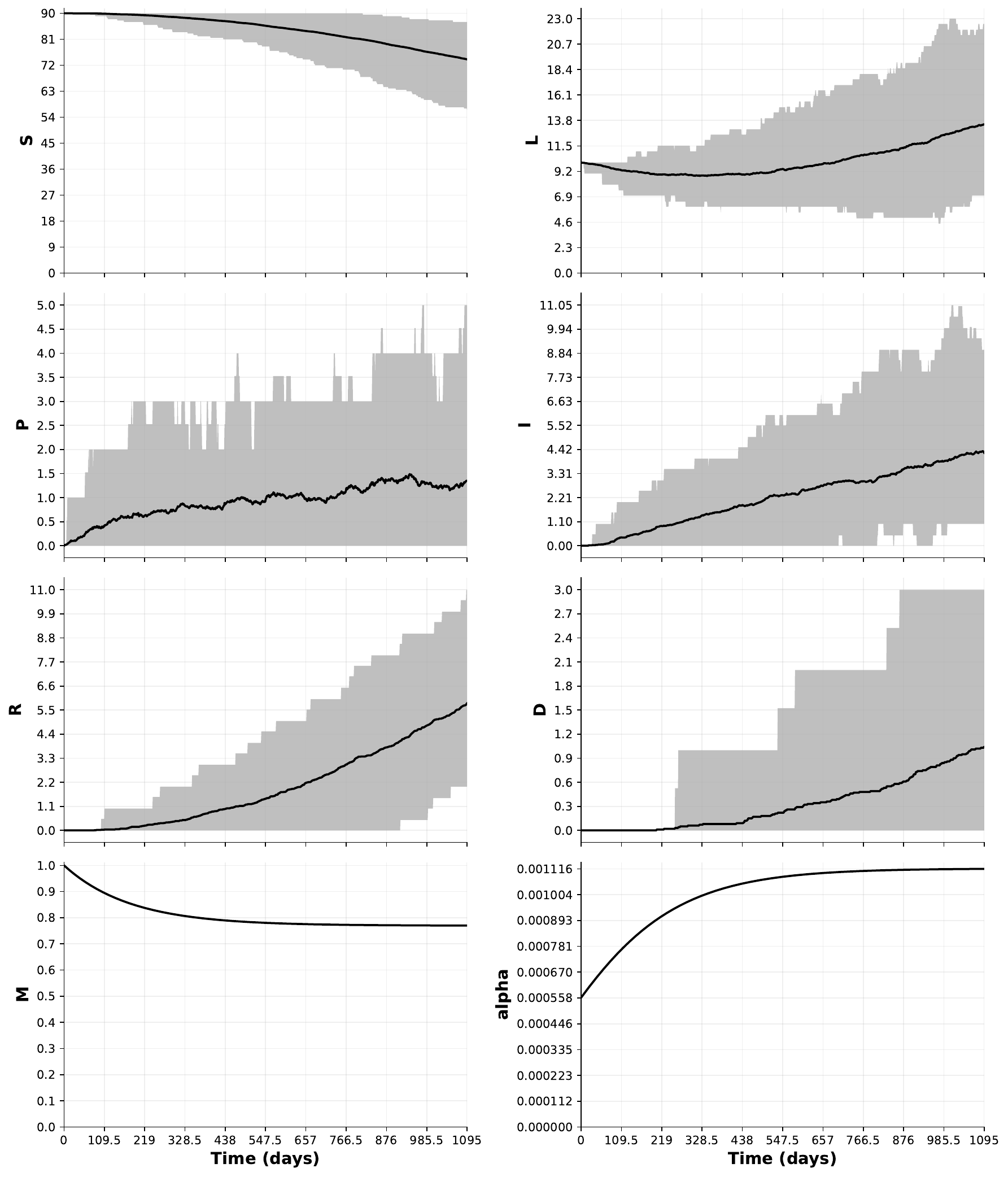}
    \caption{Uncontrolled baseline trajectories. The colony starts with latent tuberculosis carriers but no progressive or infectious active tuberculosis cases. The solid lines show mean trajectories and shaded regions indicate stochastic variability across  \(n=1000\) Monte Carlo simulations.}
    \label{fig:uncontrolled_baseline_trajectories}
\end{figure}

We first examined the natural history of latent tuberculosis reactivation in the Mars-colony model without post-reactivation intervention, i.e., with \(u_I(t)=u_M(t)=0\). The purpose of this experiment was to establish whether active tuberculosis can emerge endogenously from the latent reservoir when the colony begins with no progressive or infectious active cases. Fig.~\ref{fig:uncontrolled_baseline_trajectories} shows the baseline uncontrolled trajectories. The susceptible population decreased gradually over the three-year horizon, while latent, progressive, infectious, recovered, and dead states accumulated over time. In parallel, the colony-level immune competence \(M(t)\) declined from its initial value of 1 to approximately 0.77, while the reactivation rate \(\alpha(t)\) increased. This supports the central mechanism of the model: chronic radiation does not create infection directly, but reduces immune competence and thereby increases latent tuberculosis reactivation.

\begin{figure}[!ht]
    \centering
    \includegraphics[width=0.85\linewidth]{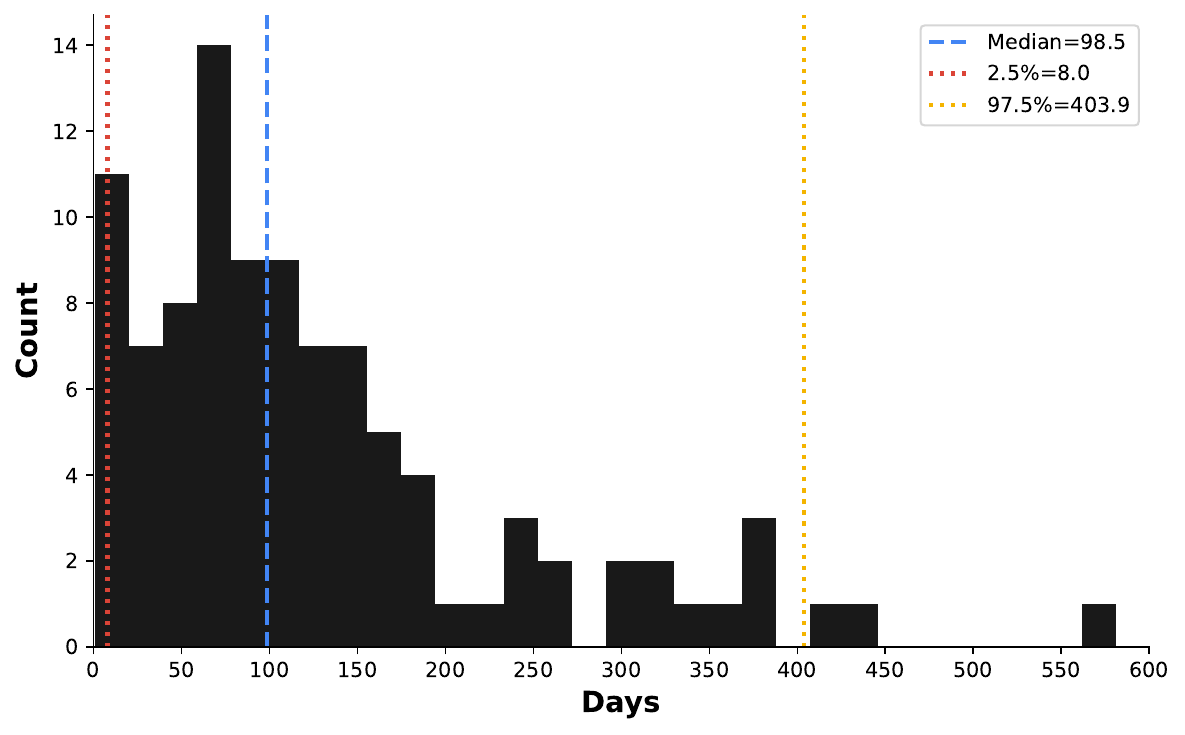}
    \caption{Distribution of the time to the first active tuberculosis case under the uncontrolled baseline scenario. The dashed vertical line marks the median and the dotted vertical lines mark the 95\% interval.}
    \label{fig:time_to_first_active_case}
\end{figure}

In the same settings, the timing of the first active case was highly stochastic. Fig.~\ref{fig:time_to_first_active_case} shows that the median time to first active tuberculosis was 98.5 days, with a wide 95\% interval of 7.95--403.9 days. Thus, even under identical baseline parameters, outbreak initiation can occur very early in some simulations and much later in others. This result is important for mission planning because the operational risk is not only the size of an outbreak after it begins, but also the uncertainty in when the first active case appears.

We next compared the learned PPO intervention policy with the four reference policies. Notably, the policies were evaluated under identical stochastic simulation settings and compared using cumulative active incidence, infectious burden, mortality, peak infectious burden, countermeasure burden, and the overall objective. Table~\ref{tab:policy_comparison_compact} shows that all intervention policies reduced disease burden relative to the no-intervention baseline, but they did so through different tradeoffs. Without intervention, the colony experienced the highest infectious burden, with a mean value of 2405 infectious person-days, 1.04 deaths, and an overall objective value of 2457. Isolation alone reduced cumulative active incidence from 12.47 to 8.22 and reduced infectious burden by 23.24\%, indicating that transmission suppression is effective once infectious cases are detected. However, its effect on mortality was moderate, with deaths reduced by 25.00\%. In particular, medication-only produced a different pattern. Although it reduced cumulative active incidence less than isolation-only, it achieved a substantially larger reduction in infectious burden and mortality. Mean deaths decreased from 1.04 under no intervention to 0.22 under medication-only, corresponding to a 78.85\% reduction. This reflects the direct effect of medication on shortening infectious duration and increasing recovery probability. The combined fixed policy achieved the strongest disease suppression overall, producing the lowest cumulative active incidence, infectious burden, peak infectious burden, deaths, and objective value. However, this improvement came at the cost of the highest countermeasure burden. Notably, The PPO-learned policy occupied an intermediate position between the passive and aggressive fixed strategies. It reduced deaths by 48.08\% and infectious burden by 31.87\% relative to no intervention, while using the lowest countermeasure burden among the active policies. Its objective value was lower than no intervention and isolation-only, but higher than medication-only and combined fixed intervention. Thus, the learned policy should not be interpreted as the most disease-suppressive strategy. Rather, it learned a resource-conserving adaptive response that provides meaningful reductions in infectious burden and mortality while avoiding the higher operational cost of the fixed combined policy.

\begin{table}[!ht]
\centering
\caption{Policy comparison summary. Values are means across Monte Carlo simulations. Percentage reductions are computed relative to the no-intervention policy.}
\label{tab:policy_comparison_compact}
\resizebox{\linewidth}{!}{
\begin{tabular}{lcccccccc}
\hline \hline
Policy & 
Cumulative active incidence & 
Infectious burden & 
Peak infectious burden & 
Deaths & 
Countermeasure burden & 
Objective & 
Death reduction & 
Infectious-burden reduction \\
\hline \hline
No intervention & 12.47 & 2405 & 5.65 & 1.04 & 0.000 & 2457 & 0.00\% & 0.00\% \\
Isolation only & 8.22 & 1846 & 4.03 & 0.78 & 1.802 & 1887 & 25.00\% & 23.24\% \\
Medication only & 10.48 & 1414 & 3.82 & 0.22 & 1.360 & 1426 & 78.85\% & 41.21\% \\
Combined fixed & 7.89 & 1156 & 3.25 & 0.19 & 2.230 & 1168 & 81.73\% & 51.93\% \\
PPO learned & 11.00 & 1638 & 4.14 & 0.54 & 0.874 & 1666 & 48.08\% & 31.87\% \\
\hline \hline
\end{tabular}
}
\end{table}

\begin{figure}[!ht]
    \centering
    \includegraphics[width=0.99\linewidth]{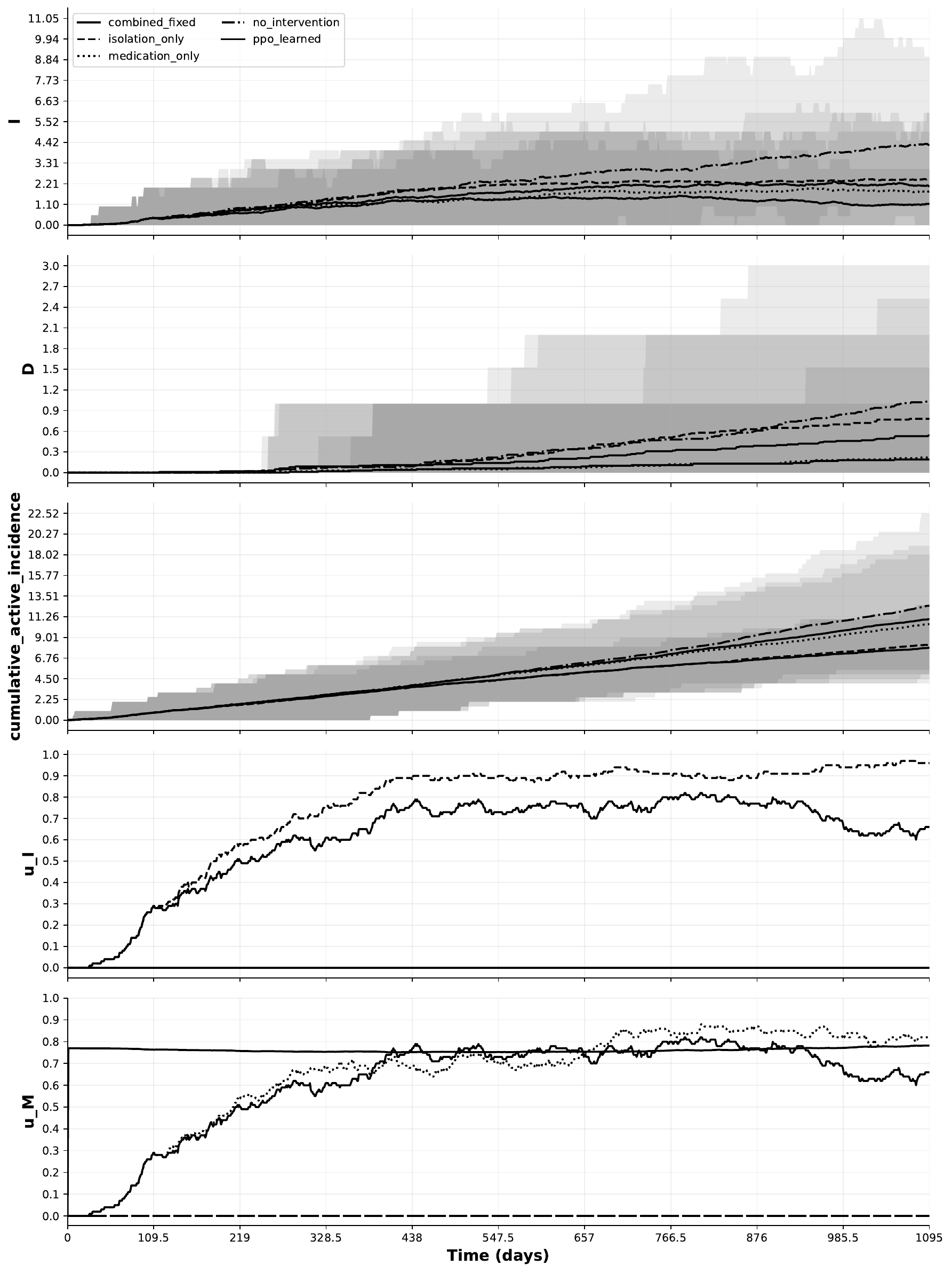}
    \caption{Policy trajectories for infectious cases, deaths, cumulative active incidence, and control actions. }
    \label{fig:policy_trajectories}
\end{figure}

The trajectory-level comparison in Fig.~\ref{fig:policy_trajectories} explains why the policies differ. Isolation-only mainly reduced transmission, whereas medication-only shortened infectious duration and improved recovery outcomes. The combined fixed policy applied both interventions after detection, producing the strongest suppression. The PPO policy, however, learned a medication-dominant strategy and used essentially no isolation in the evaluated baseline setting. This behavior reduced mortality and infectious burden while avoiding the higher intervention burden associated with fixed combined control.

We then evaluated the PPO policy under altered biological, epidemiological, radiation, and monitoring regimes. This experiment asks whether the learned controller remains useful when the underlying Mars-colony risk environment changes. Table~\ref{tab:ppo_dynamic_compact} reports representative regimes. Increasing the latent reservoir \(L(0)\), the transmission rate \(\beta\), the radiation exposure level \(G_0\), the immune-suppression coefficient \(\kappa_G\), or the reactivation sensitivity \(\theta\) increased cumulative active incidence, infectious burden, deaths, and the objective. In contrast, stronger shielding \(\eta_G\) and faster immune recovery \(\rho_M\) reduced disease burden. A notable result is that the PPO policy remained qualitatively similar across regimes. Mean isolation use remained approximately zero, whereas mean medication intensity remained around 0.75 in most settings. Thus, the learned policy generalized mainly by applying a stable medication-oriented response rather than by switching between medication and isolation. This is useful because it makes the policy interpretable, but it also reveals a limitation: under the current reward and action design, PPO did not learn a strong role for isolation.

\begin{table}[!ht]
\centering
\caption{Representative PPO outcomes under altered model regimes. Values are means across Monte Carlo simulations. The baseline regime is \(G_0=0.64\), \(L(0)=10\), \(\beta=0.0083\), \(\eta_G=0\), \(\kappa_G=0.002\), \(\rho_M=0.00556\), and \(\theta=3\).}
\label{tab:ppo_dynamic_compact}
\resizebox{\linewidth}{!}{
\begin{tabular}{llcccccc}
\hline \hline
Changed parameter & Value & 
Cumulative active incidence & 
Infectious burden & 
Deaths & 
Countermeasure burden & 
Mean \(u_I\) & 
Mean \(u_M\) \\
\hline \hline
\(G_0\) & 0.32 & 8.42 & 1223 & 0.320 & 0.655 & 0.000 & 0.763 \\
\(G_0\) & 0.64 & 10.56 & 1548 & 0.424 & 0.823 & 0.000 & 0.761 \\
\(G_0\) & 1.00 & 14.40 & 2029 & 0.604 & 1.065 & 0.000 & 0.758 \\
\hline \hline
\(L(0)\) & 1 & 1.14 & 171.2 & 0.020 & 0.093 & 0.000 & 0.768 \\
\(L(0)\) & 10 & 10.56 & 1548 & 0.424 & 0.823 & 0.000 & 0.761 \\
\(L(0)\) & 20 & 20.33 & 3061 & 0.900 & 1.594 & 0.000 & 0.752 \\
\hline \hline
\(\beta\) & 0.004 & 8.46 & 1323 & 0.368 & 0.710 & 0.000 & 0.763 \\
\(\beta\) & 0.0083 & 10.56 & 1548 & 0.424 & 0.823 & 0.000 & 0.761 \\
\(\beta\) & 0.016 & 15.65 & 2069 & 0.688 & 1.069 & 0.000 & 0.755 \\
\hline \hline
\(\eta_G\) & 0.00 & 10.56 & 1548 & 0.424 & 0.823 & 0.000 & 0.761 \\
\(\eta_G\) & 0.50 & 8.42 & 1223 & 0.320 & 0.655 & 0.000 & 0.763 \\
\(\eta_G\) & 0.75 & 7.49 & 1071 & 0.316 & 0.576 & 0.000 & 0.765 \\
\hline \hline
\(\kappa_G\) & 0.001 & 8.42 & 1223 & 0.320 & 0.655 & 0.000 & 0.763 \\
\(\kappa_G\) & 0.002 & 10.56 & 1548 & 0.424 & 0.823 & 0.000 & 0.761 \\
\(\kappa_G\) & 0.004 & 17.33 & 2419 & 0.764 & 1.254 & 0.000 & 0.755 \\
\hline \hline
\(\rho_M\) & 0.00278 & 15.39 & 2116 & 0.656 & 1.100 & 0.000 & 0.756 \\
\(\rho_M\) & 0.00556 & 10.54 & 1545 & 0.424 & 0.822 & 0.000 & 0.761 \\
\(\rho_M\) & 0.01111 & 8.59 & 1265 & 0.356 & 0.678 & 0.000 & 0.763 \\
\hline \hline
\(\theta\) & 1 & 7.89 & 1118 & 0.292 & 0.601 & 0.000 & 0.765 \\
\(\theta\) & 3 & 10.56 & 1548 & 0.424 & 0.823 & 0.000 & 0.761 \\
\(\theta\) & 7 & 20.55 & 2885 & 0.892 & 1.476 & 0.000 & 0.751 \\
\hline \hline
\end{tabular}
}
\end{table}

\begin{figure}[!p]
    \centering

    \begin{subfigure}{0.48\linewidth}
        \centering
        \includegraphics[width=\linewidth]{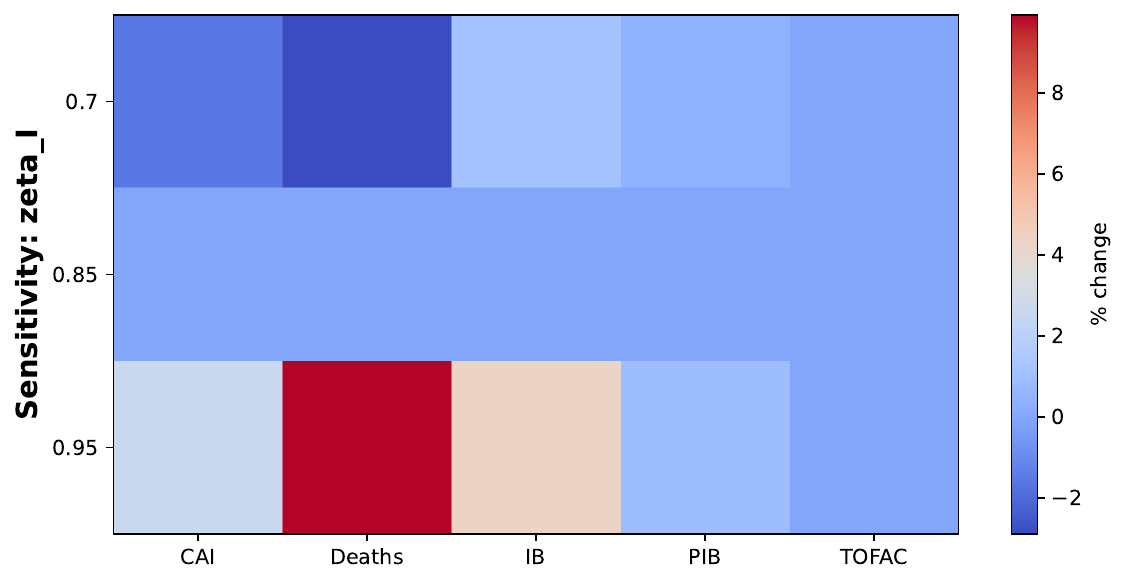}
        \caption{Infectious-state detection parameter \(\zeta_I\).}
        \label{fig:sensitivity_zeta_I}
    \end{subfigure}
    \hfill
    \begin{subfigure}{0.48\linewidth}
        \centering
        \includegraphics[width=\linewidth]{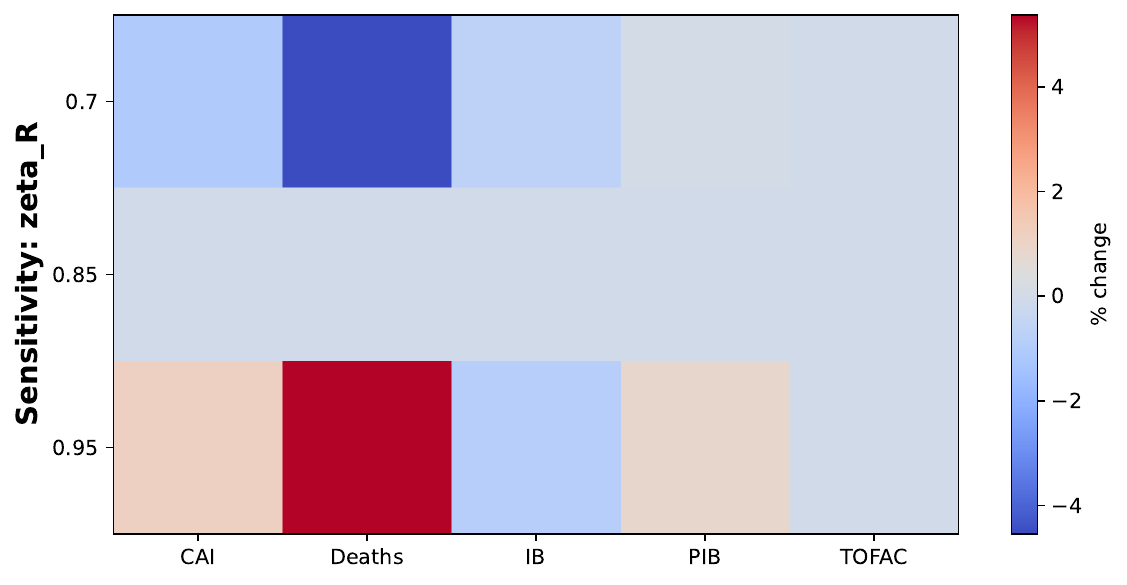}
        \caption{Recovered-state detection parameter \(\zeta_R\).}
        \label{fig:sensitivity_zeta_R}
    \end{subfigure}

    \vspace{0.35cm}

    \begin{subfigure}{0.48\linewidth}
        \centering
        \includegraphics[width=\linewidth]{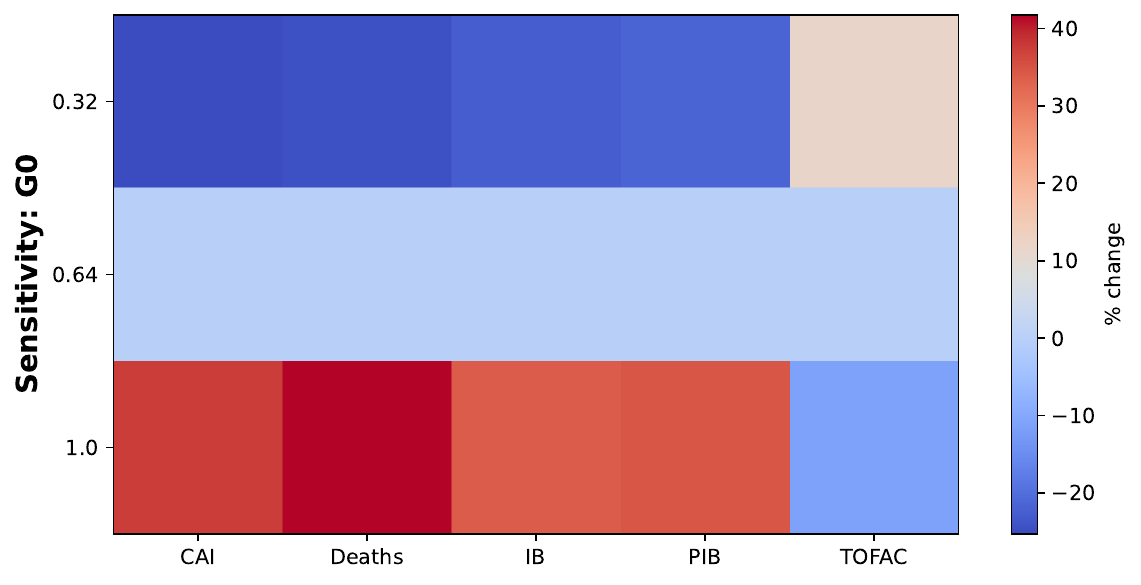}
        \caption{Chronic radiation exposure \(G_0\).}
        \label{fig:sensitivity_G0}
    \end{subfigure}
    \hfill
    \begin{subfigure}{0.48\linewidth}
        \centering
        \includegraphics[width=\linewidth]{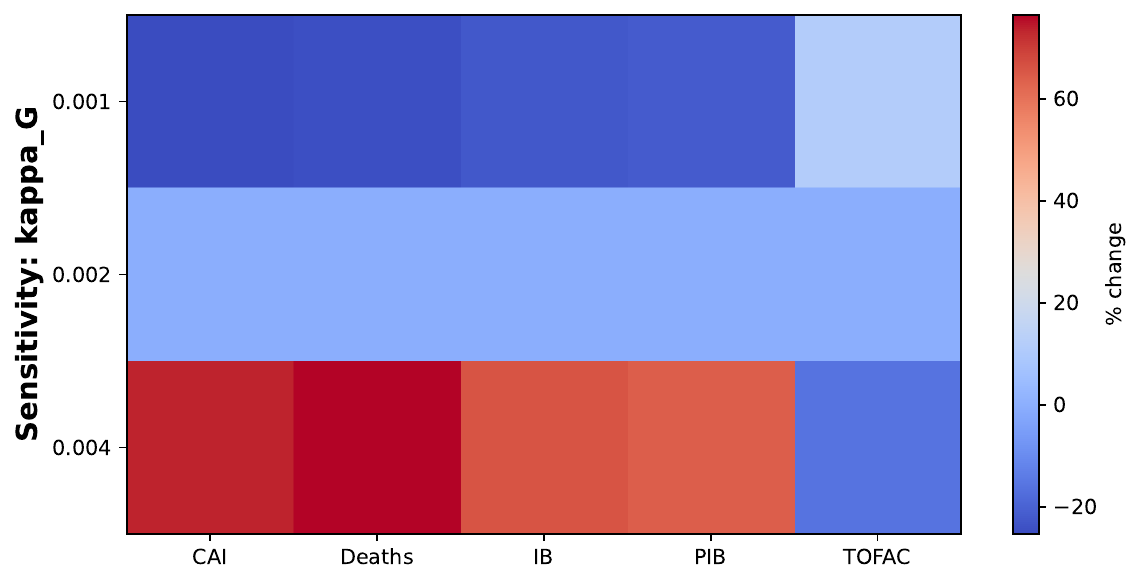}
        \caption{Radiation immune-suppression coefficient \(\kappa_G\).}
        \label{fig:sensitivity_kappa_G}
    \end{subfigure}

    \vspace{0.35cm}

    \begin{subfigure}{0.48\linewidth}
        \centering
        \includegraphics[width=\linewidth]{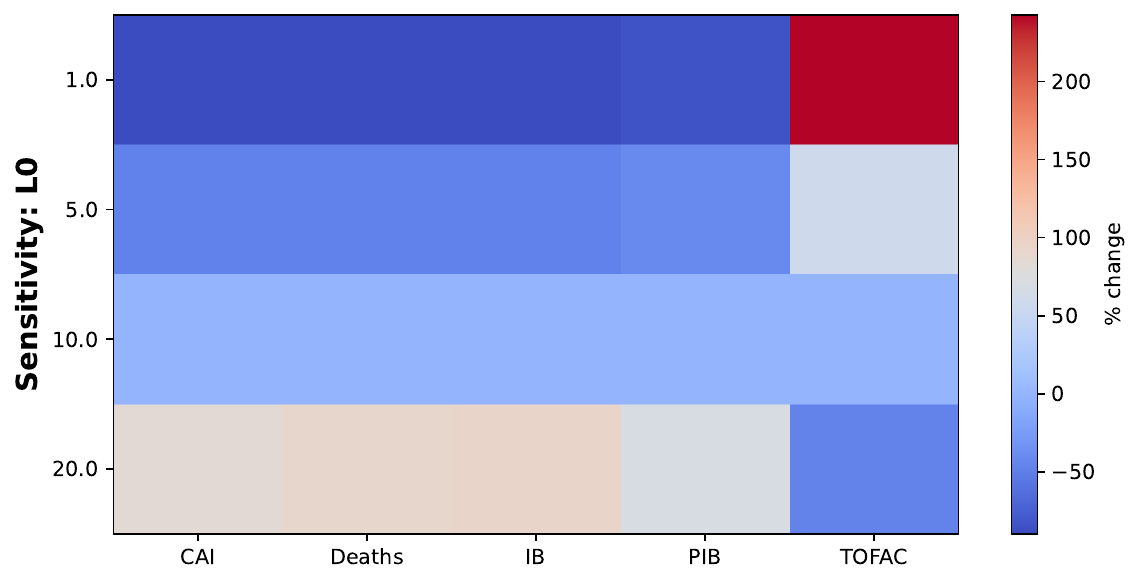}
        \caption{Initial latent reservoir \(L(0)\).}
        \label{fig:sensitivity_L0}
    \end{subfigure}
    \hfill
    \begin{subfigure}{0.48\linewidth}
        \centering
        \includegraphics[width=\linewidth]{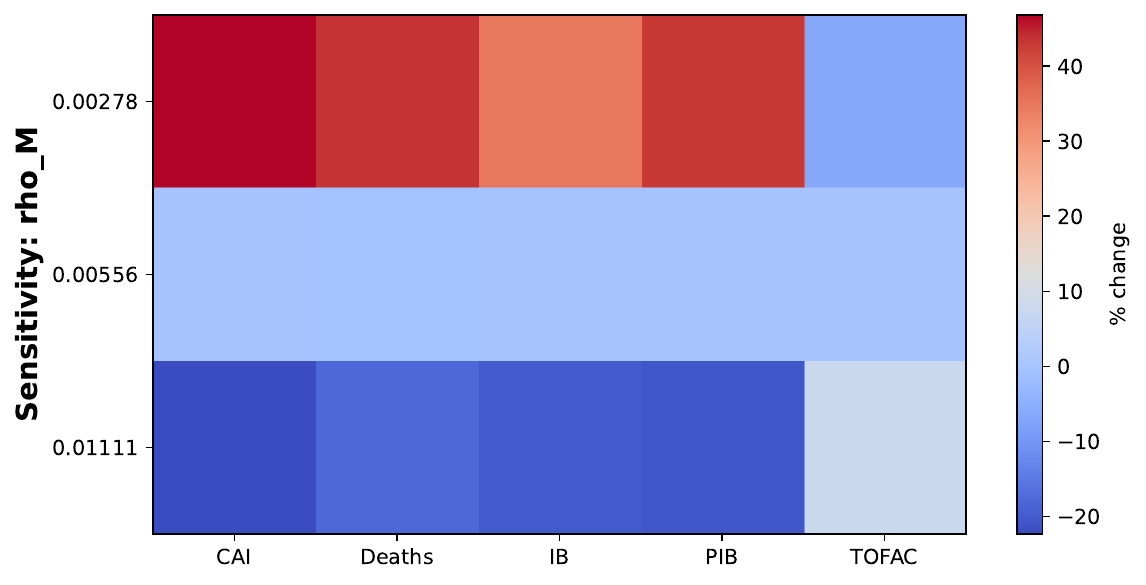}
        \caption{Immune recovery rate \(\rho_M\).}
        \label{fig:sensitivity_rho_M}
    \end{subfigure}

    \vspace{0.35cm}

    \begin{subfigure}{0.48\linewidth}
        \centering
        \includegraphics[width=\linewidth]{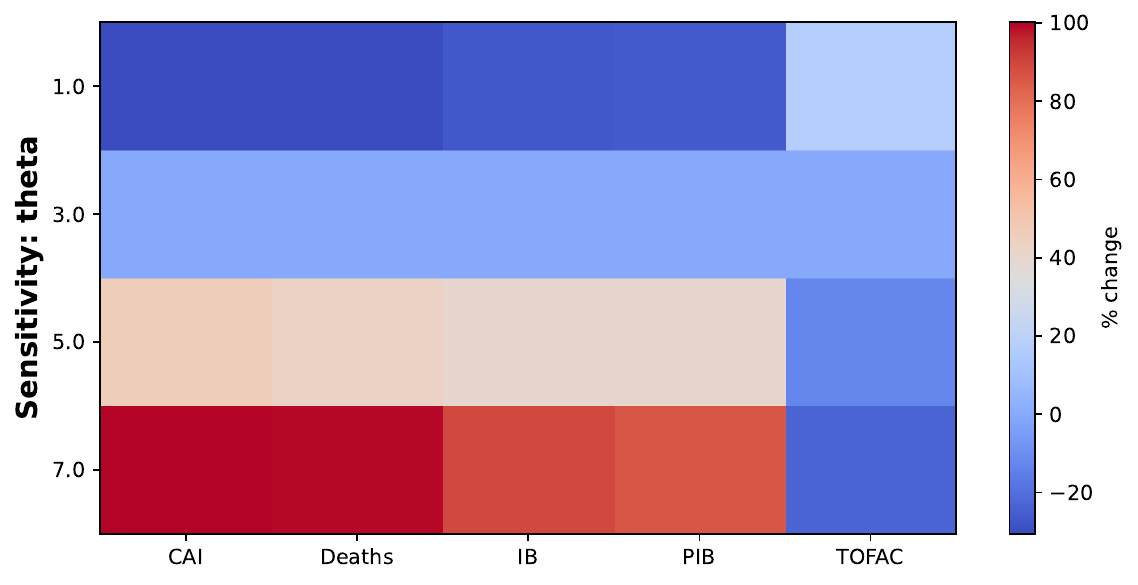}
        \caption{Reactivation sensitivity \(\theta\).}
        \label{fig:sensitivity_theta}
    \end{subfigure}

    \caption{One-at-a-time sensitivity heatmaps. Color indicates the percentage change relative to the baseline scenario. Outcome abbreviations are cumulative active incidence (CAI), deaths, infectious burden (IB), peak infectious burden (PIB), and time to first active case (TOFAC). Red indicates an increase relative to baseline and blue indicates a decrease. }
    \label{fig:sensitivity_heatmaps}
\end{figure}

Finally, we performed a one-at-a-time sensitivity analysis to identify which model assumptions most strongly determine outbreak risk. Fig.~\ref{fig:sensitivity_heatmaps} summarizes the percentage change, relative to the baseline scenario, in five outcome metrics: cumulative active incidence (CAI), deaths, infectious burden (IB), peak infectious burden (PIB), and time to first active case (TOFAC). For CAI, deaths, IB, and PIB, positive values indicate an increase in epidemiological burden. For TOFAC, the interpretation is reversed: positive values indicate that the first active case occurs later, whereas negative values indicate earlier outbreak initiation. The sensitivity analysis shows that the strongest drivers of outbreak risk are the initial latent reservoir and the radiation--immune--reactivation pathway. The initial number of latent carriers \(L(0)\) has a particularly large effect on TOFAC. When \(L(0)\) is small, the first active case is substantially delayed, whereas larger latent reservoirs shorten the time to first active disease and increase cumulative active incidence, deaths, and infectious burden. This result indicates that pre-mission latent tuberculosis screening is one of the most important prevention measures in the model, because it affects the probability and timing of endogenous reactivation before any transmission chain begins. In addition, the radiation-related parameters also have a consistent effect. Increasing the chronic radiation exposure \(G_0\) increases CAI, deaths, IB, and PIB, while reducing TOFAC. A similar pattern appears for the radiation immune-suppression coefficient \(\kappa_G\). Thus, stronger radiation exposure or stronger immune sensitivity to radiation leads to earlier and larger outbreaks. The reactivation sensitivity parameter \(\theta\) produces one of the clearest monotonic effects: larger values sharply increase disease burden and shift the first active case earlier. In contrast, faster immune recovery, represented by larger \(\rho_M\), reduces most disease-burden metrics. This supports the central mechanism of the model: outbreak initiation is governed not only by tuberculosis transmission, but by the interaction between radiation exposure, immune suppression, and latent tuberculosis reactivation. Furthermore, the monitoring parameters \(\zeta_I\) and \(\zeta_R\) show weaker effects than the biological and radiation parameters. Since larger \(\zeta\) values correspond to a higher probability of remaining undetected at each time step, increasing \(\zeta_I\) modestly worsens several outcomes, especially deaths. This occurs because delayed infectious-case detection delays the application of countermeasures. The effect of \(\zeta_R\) is smaller and less systematic, as recovered-state detection has a weaker direct influence on transmission and treatment decisions than infectious-state detection. Overall, the sensitivity results separate two mechanisms: the timing of outbreak initiation is controlled mainly by the latent reservoir and radiation-mediated reactivation, whereas the magnitude of the outbreak after initiation is shaped by the resulting infectious burden and the ability to detect and treat active cases.

\section{Discussion}
In this study, we developed a stochastic host--radiation--pathogen--habitat model to examine how chronic Mars-surface radiation may contribute to latent tuberculosis reactivation through immune suppression, and how countermeasures can be allocated once active disease appears in the colony. The main outcome is that latent tuberculosis reactivation can become an operationally meaningful risk even when the colony begins with no active infectious cases. Across the experiments, the uncontrolled simulations showed that the first active case is typically driven by the interaction between latent infection, immune suppression, and reactivation sensitivity. In contrast, the intervention experiments showed that adaptive control can reduce infectious burden and deaths more efficiently than fixed countermeasure rules, with the PPO-learned policy providing the best balance between disease suppression and countermeasure burden.

The uncontrolled simulations indicate that the most important early event is not a large transmission chain, but the first transition from latent infection into progressive active disease. Fig.~\ref{fig:time_to_first_active_case} shows that the median time to first active tuberculosis was 98.5 days, but with a wide 95\% interval. This wide distribution is important because it means that two colonies with identical parameters may face very different operational timelines. In this respect, the results agree with broader stochastic epidemic theory: early outbreak behavior in small or structured populations can be dominated by chance events, so mean trajectories alone are insufficient for risk planning \cite{lloydsmith2005superspreading}. For a Mars colony, this implies that health planning should be based not only on expected outbreak size, but also on early-reactivation scenarios and high-percentile medical-resource demand. The intervention results further show that the choice of control policy changes both epidemiological burden and operational burden. As summarized in Table~\ref{tab:policy_comparison_compact}, all active policies reduced infectious burden relative to no intervention, but no single policy dominated all metrics. The combined fixed policy produced the lowest cumulative active incidence, infectious burden, peak infectious burden, deaths, and objective value, but it also had the highest countermeasure burden, as expected. The dynamic-regime analysis in Table~\ref{tab:ppo_dynamic_compact} reinforces this interpretation. Across changes in \(G_0\), \(L(0)\), \(\beta\), \(\eta_G\), \(\kappa_G\), \(\rho_M\), and \(\theta\), the learned policy retained nearly the same qualitative action pattern: mean medication intensity remained high, whereas mean isolation remained approximately zero. This makes the policy interpretable, but it also shows that the policy did not learn a rich switching strategy between isolation and medication. The result refines the reinforcement-learning contribution of the paper: PPO is useful here as a partially observable adaptive policy benchmark, but the learned behavior remains shaped by reward design, intervention costs, and the available observation variables. This agrees with broader work on real-world reinforcement learning, which emphasizes that deployment-relevant RL must address partial observability, safety, robustness, and reward specification rather than only maximizing simulated return \cite{dulacarnold2021challenges,crucian2018immune}.

From an applicative perspective, the main lesson of the model is that the colony should not wait for a large number of infectious cases before acting aggressively and allocating a lot of resources. In many simulations, the operationally important event was not a large transmission chain but the first radiation-mediated transition from latent infection to progressive active disease. This means that a Mars-colony health system should treat the appearance of even a single detected infectious case as evidence that an unobserved reactivation process may already be underway. A useful operational rule emerging from the model is therefore to respond to the first confirmed or strongly suspected active case with a short, high-intensity diagnostic and containment phase, rather than with a gradual escalation policy designed for large terrestrial populations. In a small colony, the difference between one and two infectious cases may already represent a major increase in medical workload, contact tracing burden, and mission disruption.

A second implication is that resources should be allocated to reduce uncertainty, not only to treat visible disease. The PPO policy performed well because it was trained under partial observability: it could not see \(L(t)\), \(P(t)\), or \(M(t)\), and therefore had to infer hidden risk from delayed and imperfect observations. This suggests that monitoring quality is itself a countermeasure. Improving the probability and speed of detecting active tuberculosis may be as valuable as increasing medication or isolation capacity, because earlier detection changes the timing of intervention before the infectious burden accumulates. Practically, this supports investing in repeated respiratory screening, symptom-triggered molecular testing, environmental air monitoring, and decision rules that account for detection delay. 

A third lesson is that radiation shielding and medical stockpiles should be planned jointly. The sensitivity analysis indicates that shielding does not only affect classical radiation-health outcomes; by preserving immune competence, it can also change the expected timing and probability of tuberculosis reactivation. Therefore, shielding scenarios should be evaluated using infection outcomes such as time to first active case, cumulative active incidence, and peak infectious burden, not only dose-based health endpoints. Similarly, medication and isolation capacity should not be sized according to the mean outbreak trajectory alone. Because reactivation is stochastic, rare early-reactivation episodes can create disproportionate demand. A more useful planning output from this model is the upper-tail requirement: the amount of medication, isolation time, and diagnostic workload needed to cover high-percentile outbreak trajectories.

Despite its usability, this study is not without limitations. First, the model uses simplified immune dynamics summarized by a single colony-level immune competence variable, whereas real immune responses differ across individuals and involve interacting cellular, molecular, circadian, stress, and microbiome pathways \cite{winer2026astroimmunology,kim2024singlecell}. Second, several parameters are assumed or calibrated indirectly because there are no direct empirical data for tuberculosis reactivation under Mars-colony radiation, confinement, and monitoring conditions; indeed, NASA identifies the clinical significance of exploration-class immune dysregulation as an open gap, and recent medical-astrobiology reviews emphasize the scarcity of mission-specific epidemiological and diagnostic data \cite{nasa2026hrrimmune,boschert2025spaceflight}. Third, the well-mixed contact assumption ignores spatial habitat structure, crew schedules, ventilation zones, and task-based contact networks, all of which may strongly shape airborne tuberculosis transmission in enclosed settings \cite{nardell2016tbTransmission,patterson2022ventilation}. Fourth, the PPO policy is trained and evaluated inside the same simulated model family, so its apparent advantage may partly reflect model-specific reward and observation assumptions, a known concern in reinforcement learning where policies can exploit proxy rewards or simulation-specific design choices \cite{pan2022rewardMisspecification,bolshov2026rlReview}. Fifth, the intervention model abstracts medication into a single intensity variable and does not represent drug resistance, side effects, adherence, diagnostic confirmation, or detailed treatment regimens, whereas current tuberculosis treatment guidance distinguishes drug-susceptible and drug-resistant disease and emphasizes regimen-specific treatment and care decisions \cite{who2025tbtreatmentcare,who2022drtb}.

Taken jointly, the results show that latent tuberculosis in a Mars colony should be understood as a coupled host--pathogen--environment control problem rather than as a conventional terrestrial outbreak transplanted into space. The proposed model demonstrates how radiation-mediated immune suppression can convert a silent latent reservoir into an active operational hazard, while also showing that adaptive countermeasure allocation can reduce disease impact under partial observability. Although the numerical findings should be interpreted as scenario-based rather than predictive, the framework provides a useful foundation for comparing prevention strategies, designing autonomous health-support tools, and integrating infectious-disease risk into broader Mars-mission medical planning.

\section*{Declarations}
\subsection*{Funding}
This study received no funding. 

\subsection*{Conflicts of interest/Competing interests}
None.

\subsection*{Code and Data availability}
The code and data is provided as a supplementary material to this study.

\subsection*{Acknowledgments}
The author wishes to thank Charlotte Gaia Lazebnik for inspiring this research and Mary the cat for the support during the writing of the manuscript.

\bibliography{biblio}

@article{afshinnekoo2020fundamental,
  author  = {Afshinnekoo, Ebrahim and Scott, Ryan T. and MacKay, Matthew J. and Pariset, Eloise and Cekanaviciute, Egle and Barker, Richard and Gilroy, Simon and Hassane, Duane and Smith, Scott M. and Zwart, Sara R. and others},
  title   = {Fundamental Biological Features of Spaceflight: Advancing the Field to Enable Deep-Space Exploration},
  journal = {Cell},
  volume  = {183},
  number  = {5},
  pages   = {1162--1184.e1},
  year    = {2020},
  doi     = {10.1016/j.cell.2020.10.050}
}

@article{cowen2024infections,
  author  = {Cowen, Daniel and Zhang, Rulan and Komorowski, Matthieu},
  title   = {Infections in Long-Duration Space Missions},
  journal = {The Lancet Microbe},
  volume  = {5},
  number  = {9},
  year    = {2024},
  doi     = {10.1016/S2666-5247(24)00098-3}
}

@article{tran2021evidence,
  author  = {Tran, Kim-Anh and Pollock, Neal William and Rh{\'e}aume, Caroline and Razdan, Payal Sonya and Fortier, F{\'e}lix-Antoine and Dutil-Fafard, Lara and Morin, Camille and Monnot, David Pierre-Marie and Huot-Lavoie, Maxime and Simard-Sauriol, Philippe and others},
  title   = {Evidence Supporting the Management of Medical Conditions During Long-Duration Spaceflight: Protocol for a Scoping Review},
  journal = {JMIR Research Protocols},
  volume  = {10},
  number  = {3},
  pages   = {e24323},
  year    = {2021},
  doi     = {10.2196/24323}
}

@article{mora2016confined,
  author  = {Mora, Maximilian and Mahnert, Alexander and Koskinen, Karin and Pausan, Michelle-Ruth and Oberauner-Wappis, Lisa and Krause, Regina and Perras, Alexandra K. and Gorkiewicz, Gregor and Berg, Gabriele and Moissl-Eichinger, Christine},
  title   = {Microorganisms in Confined Habitats: Microbial Monitoring and Control of Intensive Care Units, Operating Rooms, Cleanrooms and the International Space Station},
  journal = {Frontiers in Microbiology},
  volume  = {7},
  pages   = {1573},
  year    = {2016},
  doi     = {10.3389/fmicb.2016.01573}
}

@article{slaba2025validated,
  author  = {Slaba, Tony C. and Werneth, Charles M. and Zeitlin, Cary and others},
  title   = {Validated Space Radiation Exposure Predictions from Earth to Mars During Artemis-I},
  journal = {npj Microgravity},
  volume  = {11},
  pages   = {6},
  year    = {2025},
  doi     = {10.1038/s41526-025-00459-y}
}

@article{hassler2014mars,
  author  = {Hassler, Donald M. and Zeitlin, Cary and Wimmer-Schweingruber, Robert F. and Ehresmann, Bent and Rafkin, Scot and Eigenbrode, Jennifer L. and Brinza, David E. and Weigle, G{\"u}nther and B{\"o}ttcher, Stephan and B{\"o}hm, Eckart and others},
  title   = {Mars' Surface Radiation Environment Measured with the Mars Science Laboratory's Curiosity Rover},
  journal = {Science},
  volume  = {343},
  number  = {6169},
  pages   = {1244797},
  year    = {2014},
  doi     = {10.1126/science.1244797}
}

@article{zeitlin2013measurements,
  author  = {Zeitlin, Cary and Hassler, Donald M. and Cucinotta, Francis A. and Ehresmann, Bent and Wimmer-Schweingruber, Robert F. and Brinza, David E. and Kang, Sang Y. and Weigle, G{\"u}nther and B{\"o}ttcher, Stephan and B{\"o}hm, Eckart and others},
  title   = {Measurements of Energetic Particle Radiation in Transit to Mars on the Mars Science Laboratory},
  journal = {Science},
  volume  = {340},
  number  = {6136},
  pages   = {1080--1084},
  year    = {2013},
  doi     = {10.1126/science.1235989}
}

@article{winer2026astroimmunology,
  author  = {Winer, Daniel A. and Du, Hongying and Kim, JangKeun and others},
  title   = {Astroimmunology: The Effects of Spaceflight and Its Associated Stressors on the Immune System},
  journal = {Nature Reviews Immunology},
  volume  = {26},
  pages   = {189--212},
  year    = {2026},
  doi     = {10.1038/s41577-025-01226-6}
}

@misc{nasa2024microbiology,
  author       = {{NASA Office of the Chief Health and Medical Officer}},
  title        = {{NASA-STD-3001 Technical Brief: Microbiology in Space Overview}},
  year         = {2024},
  howpublished = {Technical brief},
  institution  = {National Aeronautics and Space Administration},
  url          = {https://www.nasa.gov/wp-content/uploads/2024/05/microbiology-in-space-overview.pdf}
}

@misc{nasa2026hrrimmune,
  author       = {{NASA Human Research Program}},
  title        = {Spaceflight-Induced Immune System Dysregulation and Microgravity-Associated Alterations in Microbial Virulence -- Infectious Disease Risk for Astronauts},
  year         = {2026},
  howpublished = {Human Research Roadmap task page},
  url          = {https://humanresearchroadmap.nasa.gov/Tasks/task.aspx?i=2543}
}

@article{blue2026hsp,
  author  = {Blue, Rebecca S. and Mulcahy, Robert A. and Kreykes, Amy J. and Haddon, Robert and Pattarini, James M. and Johansen, Benjamin D. and Suresh, Rahul},
  title   = {Infectious Disease Outcomes of NASA's Health Stabilization Program},
  journal = {npj Microgravity},
  year    = {2026},
  doi     = {10.1038/s41526-026-00593-1}
}

@article{antonsen2022estimating,
  author  = {Antonsen, Erik L. and Myers, Jerry G. and Boley, Lynn and Arellano, John and Kerstman, Eric and Kadwa, Binaifer and Buckland, Daniel M. and Van Baalen, Mary},
  title   = {Estimating Medical Risk in Human Spaceflight},
  journal = {npj Microgravity},
  volume  = {8},
  number  = {1},
  year    = {2022},
  doi     = {10.1038/s41526-022-00193-9}
}

@techreport{prelich2024medprat,
  author      = {Prelich, Matthew T. and Gasiewski, Clara M. and McIntyre, Lauren and Kadwa, Binaifer and Arellano, John and Myers, Jerry G.},
  title       = {Assessment of Model Outcomes Between the Integrated Medical Model (IMM) and the Medical Extensible Dynamic Probabilistic Risk Assessment Tool (MEDPRAT)},
  institution = {National Aeronautics and Space Administration},
  number      = {NASA/TM-20240012058},
  year        = {2024},
  url         = {https://ntrs.nasa.gov/citations/20240012058}
}

@article{menzies2018progression,
  author  = {Menzies, Nicolas A. and Wolf, Emory and Connors, David and Bellerose, Meghan and Sbarra, Alyssa N. and Cohen, Ted and Hill, Andrew N. and Yaesoubi, Reza and Galer, Kara and White, Peter J. and Abubakar, Ibrahim and Salomon, Joshua A.},
  title   = {Progression from Latent Infection to Active Disease in Dynamic Tuberculosis Transmission Models: A Systematic Review of the Validity of Modelling Assumptions},
  journal = {The Lancet Infectious Diseases},
  volume  = {18},
  number  = {8},
  pages   = {e228--e238},
  year    = {2018},
  doi     = {10.1016/S1473-3099(18)30134-8}
}

@article{checinska2019isssurfaces,
  author  = {Checinska Sielaff, Aleksandra and Urbaniak, Camilla and Mohan, Ganesh Babu Malli and Stepanov, Victor G. and Tran, Quyen and Wood, Jason M. and Minich, Jeremiah and McDonald, Daniel and Mayer, Tasha and Knight, Rob and others},
  title   = {Characterization of the Total and Viable Bacterial and Fungal Communities Associated with the International Space Station Surfaces},
  journal = {Microbiome},
  volume  = {7},
  pages   = {50},
  year    = {2019},
  doi     = {10.1186/s40168-019-0666-x}
}

@article{voorhies2019impact,
  author  = {Voorhies, Alexander A. and Ott, C. Mark and Mehta, Satish and Pierson, Duane L. and Crucian, Brian E. and Feiveson, Alan and Oubre, Cherie M. and Torralba, Manolito and Moncera, Kevin and Zhang, Yi and others},
  title   = {Study of the Impact of Long-Duration Space Missions at the International Space Station on the Astronaut Microbiome},
  journal = {Scientific Reports},
  volume  = {9},
  pages   = {9911},
  year    = {2019},
  doi     = {10.1038/s41598-019-46303-8}
}

@article{mehta2017latent,
  author  = {Mehta, Satish K. and Crucian, Brian E. and Stowe, Raymond P. and Simpson, Richard J. and Ott, C. Mark and Sams, Clarence F. and Pierson, Duane L.},
  title   = {Latent Virus Reactivation in Astronauts on the International Space Station},
  journal = {npj Microgravity},
  volume  = {3},
  pages   = {11},
  year    = {2017},
  doi     = {10.1038/s41526-017-0015-y}
}

@article{rooney2019herpes,
  author  = {Rooney, Bridgette V. and Crucian, Brian E. and Pierson, Duane L. and Laudenslager, Mark L. and Mehta, Satish K.},
  title   = {Herpes Virus Reactivation in Astronauts During Spaceflight and Its Application on Earth},
  journal = {Frontiers in Microbiology},
  volume  = {10},
  pages   = {16},
  year    = {2019},
  doi     = {10.3389/fmicb.2019.00016}
}

@misc{who2018ltbi,
  author       = {{World Health Organization}},
  title        = {Latent Tuberculosis Infection: Updated and Consolidated Guidelines for Programmatic Management},
  year         = {2018},
  institution  = {World Health Organization},
  address      = {Geneva},
  isbn         = {978-92-4-155023-9},
  url          = {https://www.who.int/publications/i/item/9789241550239}
}

@misc{cdc2024latent,
  author       = {{Centers for Disease Control and Prevention}},
  title        = {Clinical Overview of Latent Tuberculosis Infection},
  year         = {2024},
  howpublished = {Web page},
  url          = {https://www.cdc.gov/tb/hcp/clinical-overview/latent-tuberculosis-infection.html}
}

@misc{cdc2025spread,
  author       = {{Centers for Disease Control and Prevention}},
  title        = {Tuberculosis: Causes and How It Spreads},
  year         = {2025},
  howpublished = {Web page},
  url          = {https://www.cdc.gov/tb/causes/index.html}
}

@article{pai2016tuberculosis,
  author  = {Pai, Madhukar and Behr, Marcel A. and Dowdy, David and Dheda, Keertan and Divangahi, Maziar and Boehme, Catharina C. and Ginsberg, Ann and Swaminathan, Soumya and Spigelman, Melvin and Getahun, Haileyesus and Menzies, Dick and Raviglione, Mario},
  title   = {Tuberculosis},
  journal = {Nature Reviews Disease Primers},
  volume  = {2},
  pages   = {16076},
  year    = {2016},
  doi     = {10.1038/nrdp.2016.76}
}

@article{behr2021reactivation,
  author  = {Behr, Marcel A. and Edelstein, Paul H. and Ramakrishnan, Lalita},
  title   = {Quantifying the Rates of Late Reactivation Tuberculosis: A Systematic Review},
  journal = {The Lancet Infectious Diseases},
  year    = {2021},
  doi     = {10.1016/S1473-3099(20)30728-3}
}

@article{rangaka2012predictive,
  author  = {Rangaka, Molebogeng X. and Wilkinson, Katherine A. and Glynn, Judith R. and Ling, Dena and Menzies, Dick and Mwansa-Kambafwile, Judith and Fielding, Katherine and Wilkinson, Robert J. and Pai, Madhukar},
  title   = {Predictive Value of Interferon-\(\gamma\) Release Assays for Incident Active Tuberculosis: A Systematic Review and Meta-Analysis},
  journal = {The Lancet Infectious Diseases},
  volume  = {12},
  number  = {1},
  pages   = {45--55},
  year    = {2012},
  doi     = {10.1016/S1473-3099(11)70210-9}
}

@article{garrettbakelman2019twins,
  author  = {Garrett-Bakelman, Francine E. and Darshi, Manjula and Green, Stefan J. and Gur, Ruben C. and Lin, Ling and Macias, Brandon R. and McKenna, Miles J. and Meydan, Cem and Mishra, Tejaswini and Nasrini, Jad and others},
  title   = {The NASA Twins Study: A Multidimensional Analysis of a Year-Long Human Spaceflight},
  journal = {Science},
  volume  = {364},
  number  = {6436},
  pages   = {eaau8650},
  year    = {2019},
  doi     = {10.1126/science.aau8650}
}

@article{tierney2024longitudinal,
  author  = {Tierney, Braden T. and Overbey, Eliah G. and others},
  title   = {Longitudinal Multi-Omics Analysis of Host Microbiome Architecture and Immune Responses During Short-Term Spaceflight},
  journal = {Nature Microbiology},
  year    = {2024},
  doi     = {10.1038/s41564-024-01635-8}
}

@article{kim2024singlecell,
  author  = {Kim, JangKeun and Tierney, Braden T. and Overbey, Eliah G. and Dantas, Ezequiel and Fuentealba, Matias and Park, Jiwoon and Narayanan, S. Anand and Wu, Fei and Najjar, Deena and Chin, Christopher R. and others},
  title   = {Single-Cell Multi-Ome and Immune Profiles of the Inspiration4 Crew Reveal Conserved, Cell-Type, and Sex-Specific Responses to Spaceflight},
  journal = {Nature Communications},
  year    = {2024},
  doi     = {10.1038/s41467-024-49211-2}
}

@article{salido2025iss,
  author  = {Salido, Rodolfo A. and Zhao, Haoqi Nina and McDonald, Daniel and Mannochio-Russo, Helena and Zuffa, Simone and Oles, Renee E. and Aron, Allegra T. and El Abiead, Yasin and Farmer, Sawyer and Gonz{\'a}lez, Antonio and others},
  title   = {The International Space Station Has a Unique and Extreme Microbial and Chemical Environment Driven by Use Patterns},
  journal = {Cell},
  volume  = {188},
  number  = {7},
  pages   = {2022--2041.e23},
  year    = {2025},
  doi     = {10.1016/j.cell.2025.01.039}
}

@article{singh2024bugandensis,
  author  = {Singh, Nitin K. and others},
  title   = {Genomic, Functional, and Metabolic Enhancements in Multidrug-Resistant \emph{Enterobacter bugandensis} Facilitating Its Persistence and Succession in the International Space Station},
  journal = {Microbiome},
  year    = {2024},
  doi     = {10.1186/s40168-024-01777-1}
}

@article{irby2024phage,
  author  = {Irby, Iris and Broddrick, Jared T.},
  title   = {Microbial Adaptation to Spaceflight Is Correlated with Bacteriophage-Encoded Functions},
  journal = {Nature Communications},
  volume  = {15},
  number  = {1},
  year    = {2024},
  doi     = {10.1038/s41467-023-42104-w}
}

@article{hardy2025standardmeasures,
  author  = {Hardy, John G. and Theriot, Corey A. and Oswald, Thomas and Cl{\'e}ment, Gilles},
  title   = {Spaceflight Standard Measures Is a Multidisciplinary Study That Systematically Monitors Risks to Astronaut Health and Performance},
  journal = {npj Microgravity},
  year    = {2025},
  doi     = {10.1038/s41526-025-00532-6}
}

@article{boschert2025spaceflight,
  author  = {Boschert, Alessa Lalinka and Leuko, Stefan and Kr{\"a}mer, Carolin Luisa and Siems, Katharina and Ly-Sauerbrey, Yen-Tran and Arndt, Franca},
  title   = {Spaceflight and Medical Microbiology: Possible Implications for Standard Infection Diagnostics and Therapy},
  journal = {Life},
  volume  = {15},
  number  = {11},
  pages   = {1757},
  year    = {2025},
  doi     = {10.3390/life15111757}
}

@book{hessel2022space,
  title={In-Space Manufacturing and Resources},
  author={Hessel, Volker and Stoudemire, Jana and Miyamoto, Hideaki and Fisk, ID},
  year={2022},
  publisher={Wiley Online Library}
}

@article{neukart2024towards,
  title={Towards sustainable horizons: A comprehensive blueprint for Mars colonization},
  author={Neukart, Florian},
  journal={Heliyon},
  volume={10},
  number={4},
  year={2024},
  publisher={Elsevier}
}

@article{castro2024infectious,
  title={Infectious diseases and the use of antimicrobials on space missions},
  author={Castro-Costa, Alice RC e and Siqueira-Batista, Rodrigo and Alc{\^a}ntara, Fab{\'\i}ola A and Russomano, Tha{\'\i}s and Santos, Marlise A and Silva, Isadora de C e and Del Cima, Oswaldo M},
  journal={Space: Science \& Technology},
  volume={4},
  pages={0205},
  year={2024},
  publisher={AAAS}
}

@article{zaccaria2025effects,
  title={Effects of simulated Martian environmental stressors on specific human pathogen--immune system interactions},
  author={Zaccaria, Tommaso and Bulut, {\"O}zlem and Ferreira, Anaisa V and Dona, Margo and Langereis, Jeroen D and Mesman, Rob J and Wesseling, Joppe and van Niftrik, Laura and Netea, Mihai G and Rettberg, Petra and others},
  journal={Mbio},
  volume={16},
  number={9},
  pages={e01099--25},
  year={2025},
  publisher={American Society for Microbiology 1752 N St., NW, Washington, DC}
}

@article{khoshtinat2025earth,
  title={From Earth to Mars: a perspective on exploiting biomineralization for Martian construction},
  author={Khoshtinat, Shiva and Long-Fox, Jared and Hosseini, Seyed Mohammad Javad},
  journal={Frontiers in Microbiology},
  volume={16},
  pages={1645014},
  year={2025},
  publisher={Frontiers Media SA}
}

@article{allen2017primer, author = {Allen, Linda J. S.}, title = {A Primer on Stochastic Epidemic Models: Formulation, Numerical Simulation, and Analysis}, journal = {Infectious Disease Modelling}, volume = {2}, number = {2}, pages = {128--142}, year = {2017}, doi = {10.1016/j.idm.2017.03.001} }

@article{kermack1927contribution,
  author  = {Kermack, William O. and McKendrick, Anderson G.},
  title   = {A Contribution to the Mathematical Theory of Epidemics},
  journal = {Proceedings of the Royal Society of London. Series A},
  volume  = {115},
  number  = {772},
  pages   = {700--721},
  year    = {1927},
  doi     = {10.1098/rspa.1927.0118}
}

@article{hethcote2000mathematics,
  author  = {Hethcote, Herbert W.},
  title   = {The Mathematics of Infectious Diseases},
  journal = {SIAM Review},
  volume  = {42},
  number  = {4},
  pages   = {599--653},
  year    = {2000},
  doi     = {10.1137/S0036144500371907}
}

@book{brauer2019mathematical,
  author    = {Brauer, Fred and Castillo-Chavez, Carlos and Feng, Zhilan},
  title     = {Mathematical Models in Epidemiology},
  publisher = {Springer},
  address   = {New York},
  year      = {2019},
  doi       = {10.1007/978-1-4939-9828-9}
}

@article{cox1972regression,
  author  = {Cox, David R.},
  title   = {Regression Models and Life-Tables},
  journal = {Journal of the Royal Statistical Society: Series B},
  volume  = {34},
  number  = {2},
  pages   = {187--202},
  year    = {1972},
  doi     = {10.1111/j.2517-6161.1972.tb00899.x}
}

@misc{cdc2023tbinfectioncontrol,
  author       = {{Centers for Disease Control and Prevention}},
  title        = {Tuberculosis Infection Control},
  year         = {2023},
  howpublished = {Web page},
  url          = {https://www.cdc.gov/tb-healthcare-settings/hcp/infection-control/index.html}
}

@misc{cdc2025dstreatment,
  author       = {{Centers for Disease Control and Prevention}},
  title        = {Treatment for Drug-Susceptible Tuberculosis Disease},
  year         = {2025},
  howpublished = {Web page},
  url          = {https://www.cdc.gov/tb/hcp/treatment/tuberculosis-disease.html}
}

@misc{who2025tbtreatmentcare,
  author       = {{World Health Organization}},
  title        = {WHO Consolidated Guidelines on Tuberculosis: Module 4: Treatment and Care},
  year         = {2025},
  institution  = {World Health Organization},
  url          = {https://www.who.int/publications/i/item/9789240107243}
}

@book{keeling2008modeling,
  author    = {Keeling, Matthew J. and Rohani, Pejman},
  title     = {Modeling Infectious Diseases in Humans and Animals},
  publisher = {Princeton University Press},
  address   = {Princeton, NJ},
  year      = {2008},
  isbn      = {9780691116174}
}

@article{sharomi2017optimal,
  author  = {Sharomi, Oluwaseun and Malik, Tufail},
  title   = {Optimal Control in Epidemiology},
  journal = {Annals of Operations Research},
  volume  = {251},
  number  = {1--2},
  pages   = {55--71},
  year    = {2017},
  doi     = {10.1007/s10479-015-1834-4}
}

@book{lenhart2007optimal,
  author    = {Lenhart, Suzanne and Workman, John T.},
  title     = {Optimal Control Applied to Biological Models},
  publisher = {Chapman and Hall/CRC},
  address   = {Boca Raton, FL},
  year      = {2007},
  doi       = {10.1201/9781420011418}
}

@article{diekmann2010construction,
  author  = {Diekmann, Odo and Heesterbeek, J. A. P. and Roberts, M. G.},
  title   = {The Construction of Next-Generation Matrices for Compartmental Epidemic Models},
  journal = {Journal of the Royal Society Interface},
  volume  = {7},
  number  = {47},
  pages   = {873--885},
  year    = {2010},
  doi     = {10.1098/rsif.2009.0386}
}

@article{bonabeau2002agentbased,
  author  = {Bonabeau, Eric},
  title   = {Agent-Based Modeling: Methods and Techniques for Simulating Human Systems},
  journal = {Proceedings of the National Academy of Sciences},
  volume  = {99},
  pages   = {7280--7287},
  year    = {2002},
  doi     = {10.1073/pnas.082080899}
}

@book{sutton2018reinforcement,
  author    = {Sutton, Richard S. and Barto, Andrew G.},
  title     = {Reinforcement Learning: An Introduction},
  edition   = {2},
  publisher = {MIT Press},
  year      = {2018},
  isbn      = {9780262039246}
}

@article{schulman2017ppo,
  author  = {Schulman, John and Wolski, Filip and Dhariwal, Prafulla and Radford, Alec and Klimov, Oleg},
  title   = {Proximal Policy Optimization Algorithms},
  journal = {arXiv preprint arXiv:1707.06347},
  year    = {2017},
  doi     = {10.48550/arXiv.1707.06347}
}

@misc{who2024tbreport,
  author       = {{World Health Organization}},
  title        = {Global Tuberculosis Report 2024},
  year         = {2024},
  institution  = {World Health Organization},
  url          = {https://www.who.int/teams/global-programme-on-tuberculosis-and-lung-health/tb-reports/global-tuberculosis-report-2024}
}

@article{ma2018quantifying,
  author  = {Ma, Y. and Horsburgh, C. R. and White, L. F. and Jenkins, H. E.},
  title   = {Quantifying TB Transmission: A Systematic Review of Reproduction Number and Serial Interval Estimates for Tuberculosis},
  journal = {Epidemiology and Infection},
  volume  = {146},
  number  = {12},
  pages   = {1478--1494},
  year    = {2018},
  doi     = {10.1017/S0950268818001760}
}

@article{kaelbling1998planning,
  author  = {Kaelbling, Leslie Pack and Littman, Michael L. and Cassandra, Anthony R.},
  title   = {Planning and Acting in Partially Observable Stochastic Domains},
  journal = {Artificial Intelligence},
  volume  = {101},
  number  = {1--2},
  pages   = {99--134},
  year    = {1998},
  doi     = {10.1016/S0004-3702(98)00023-X}
}

@article{raffin2021stablebaselines3,
  author  = {Raffin, Antonin and Hill, Ashley and Gleave, Adam and Kanervisto, Anssi and Ernestus, Maximilian and Dormann, Noah},
  title   = {Stable-Baselines3: Reliable Reinforcement Learning Implementations},
  journal = {Journal of Machine Learning Research},
  volume  = {22},
  number  = {268},
  pages   = {1--8},
  year    = {2021}
}

@article{bolshov2026rlReview,
author  = {Bolshov, Oleksandr and Chumachenko, Dmytro},
title   = {Reinforcement learning for policymaking in epidemic control: A scoping review},
journal = {PLOS One},
year    = {2026},
volume  = {21},
number  = {6},
pages   = {e0351176},
doi     = {10.1371/journal.pone.0351176},
url     = {https://doi.org/10.1371/journal.pone.0351176}
}

@manual{who2022drtb,
author = {{World Health Organization}},
title = {{WHO} consolidated guidelines on tuberculosis. Module 4: treatment -- drug-resistant tuberculosis treatment, 2022 update},
organization = {World Health Organization},
year = {2022},
address = {Geneva},
isbn = {9789240063129},
url = {https://www.who.int/publications/i/item/9789240063129}
}

@inproceedings{pan2022rewardMisspecification,
author = {Pan, Alexander and Bhatia, Kush and Steinhardt, Jacob},
title = {The Effects of Reward Misspecification: Mapping and Mitigating Misaligned Models},
booktitle = {International Conference on Learning Representations},
year = {2022},
url = {https://openreview.net/forum?id=JYtwGwIL7ye}
}

@article{patterson2022ventilation,
author = {Deol, A. K. and Shaikh, N. and Middelkoop, K. and Mohlamonyane, M. and White, R. G. and McCreesh, N.},
title = {Importance of ventilation and occupancy to {Mycobacterium tuberculosis} transmission rates in congregate settings},
journal = {BMC Public Health},
year = {2022},
volume = {22},
pages = {1772},
doi = {10.1186/s12889-022-14133-5},
url = {https://doi.org/10.1186/s12889-022-14133-5}
}

@article{nardell2016tbTransmission,
author = {Nardell, Edward A.},
title = {Transmission and Institutional Infection Control of Tuberculosis},
journal = {Cold Spring Harbor Perspectives in Medicine},
year = {2016},
volume = {6},
number = {2},
pages = {a018192},
doi = {10.1101/cshperspect.a018192},
url = {https://doi.org/10.1101/cshperspect.a018192}
}

@article{lazebnik2023extendedSIRReview,
author  = {Lazebnik, Teddy},
title   = {Computational applications of extended {SIR} models: A review focused on airborne pandemics},
journal = {Ecological Modelling},
year    = {2023},
volume  = {483},
pages   = {110422},
doi     = {10.1016/j.ecolmodel.2023.110422},
url     = {https://doi.org/10.1016/j.ecolmodel.2023.110422}
}

@article{lazebnik2023roomAirborne,
author  = {Lazebnik, Teddy and Alexi, Ariel},
title   = {High Resolution Spatio-Temporal Model for Room-Level Airborne Pandemic Spread},
journal = {Mathematics},
year    = {2023},
volume  = {11},
number  = {2},
pages   = {426},
doi     = {10.3390/math11020426},
url     = {https://doi.org/10.3390/math11020426}
}

@article{alexi2022ventilationTradeoff,
author  = {Alexi, Ariel and Rosenfeld, Ariel and Lazebnik, Teddy},
title   = {The Trade-Off between Airborne Pandemic Control and Energy Consumption Using Air Ventilation Solutions},
journal = {Sensors},
year    = {2022},
volume  = {22},
number  = {22},
pages   = {8594},
doi     = {10.3390/s22228594},
url     = {https://doi.org/10.3390/s22228594}
}

@article{lazebnik2023hospitalDRL,
author  = {Lazebnik, Teddy},
title   = {Data-driven hospitals staff and resources allocation using agent-based simulation and deep reinforcement learning},
journal = {Engineering Applications of Artificial Intelligence},
year    = {2023},
volume  = {126},
pages   = {106783},
doi     = {10.1016/j.engappai.2023.106783},
url     = {https://doi.org/10.1016/j.engappai.2023.106783}
}

@article{shuchami2025warPandemicDRL,
author  = {Shuchami, Adi and Lazebnik, Teddy},
title   = {Spatio-Temporal {SIR} Model of Pandemic Spread During Warfare with Optimal Dual-use Health Care System Administration using Deep Reinforcement Learning},
journal = {Disaster Medicine and Public Health Preparedness},
year    = {2025},
volume  = {19},
pages   = {e197},
doi     = {10.1017/dmp.2025.10062},
url     = {https://doi.org/10.1017/dmp.2025.10062}
}

@article{lazebnik2022npiEconomy,
author  = {Lazebnik, Teddy and Shami, Labib and Bunimovich-Mendrazitsky, Svetlana},
title   = {Spatio-Temporal influence of Non-Pharmaceutical interventions policies on pandemic dynamics and the economy: the case of {COVID-19}},
journal = {Economic Research-Ekonomska Istrazivanja},
year    = {2022},
volume  = {35},
number  = {1},
pages   = {1833--1861},
doi     = {10.1080/1331677X.2021.1925573},
url     = {https://doi.org/10.1080/1331677X.2021.1925573}
}

@article{lazebnik2023borderClosure,
author  = {Lazebnik, Teddy and Shami, Labib and Bunimovich-Mendrazitsky, Svetlana},
title   = {A Hybrid Mathematical Model for an Optimal Border Closure Policy During a Pandemic},
journal = {International Journal of Applied Mathematics and Computer Science},
year    = {2023},
volume  = {33},
number  = {4},
pages   = {583--601},
doi     = {10.34768/amcs-2023-0042},
url     = {https://doi.org/10.34768/amcs-2023-0042}
}

@article{lazebnik2023pandemicManagement,
author  = {Lazebnik, Teddy and Bunimovich-Mendrazitsky, Svetlana and Shami, Labib},
title   = {Pandemic management by a spatio-temporal mathematical model},
journal = {International Journal of Nonlinear Sciences and Numerical Simulation},
year    = {2023},
volume  = {24},
number  = {6},
pages   = {2307--2324},
doi     = {10.1515/ijnsns-2021-0063},
url     = {https://doi.org/10.1515/ijnsns-2021-0063}
}

@article{alexi2023securityGames,
author  = {Alexi, Ariel and Rosenfeld, Ariel and Lazebnik, Teddy},
title   = {A Security Games Inspired Approach for Distributed Control Of Pandemic Spread},
journal = {Advanced Theory and Simulations},
year    = {2023},
volume  = {6},
number  = {2},
pages   = {2200631},
doi     = {10.1002/adts.202200631},
url     = {https://doi.org/10.1002/adts.202200631}
}

@article{lloydsmith2005superspreading,
  title={Superspreading and the effect of individual variation on disease emergence},
  author={Lloyd-Smith, James O. and Schreiber, Sebastian J. and Kopp, P. Ekkehard and Getz, Wayne M.},
  journal={Nature},
  volume={438},
  number={7066},
  pages={355--359},
  year={2005},
  doi={10.1038/nature04153}
}

@article{crucian2018immune,
  title={Immune system dysregulation during spaceflight: Potential countermeasures for deep space exploration missions},
  author={Crucian, Brian E. and Chouk{\`e}r, Alexander and Simpson, Richard J. and Mehta, Satish and Marshall, Gail and Smith, Scott M. and Zwart, Sara R. and Heer, Martina and Ponomarev, Sergey and Whitmire, Alexandra and Frippiat, Jean-Pol and Douglas, Grace L. and Lorenzi, Hernan and Buchheim, Judith-Irina and Makedonas, Georgios and Ginsburg, Geoffrey S. and Ott, C. Mark},
  journal={Frontiers in Immunology},
  volume={9},
  pages={1437},
  year={2018},
  doi={10.3389/fimmu.2018.01437}
}

@article{dulacarnold2021challenges,
  title={Challenges of real-world reinforcement learning: Definitions, benchmarks and analysis},
  author={Dulac-Arnold, Gabriel and Mankowitz, Daniel and Hester, Todd},
  journal={Machine Learning},
  volume={110},
  pages={2419--2468},
  year={2021},
  doi={10.1007/s10994-021-05961-4}
}

@article{kennedy2014biological,
  title   = {Biological effects of space radiation and development of effective countermeasures},
  author  = {Kennedy, Ann R.},
  journal = {Life Sciences in Space Research},
  volume  = {1},
  pages   = {10--43},
  year    = {2014},
  doi     = {10.1016/j.lssr.2014.02.004}
}

@article{sanzari2015leukocyte,
  title   = {Comparison of changes over time in leukocyte counts in Yucatan minipigs irradiated with simulated solar particle event-like radiation},
  author  = {Sanzari, Jenine K. and Wan, X. Steven and Muehlmatt, Amy and Lin, Liyong and Kennedy, Ann R.},
  journal = {Life Sciences in Space Research},
  volume  = {4},
  pages   = {11--16},
  year    = {2015},
  doi     = {10.1016/j.lssr.2014.12.002}
}

@article{townsend2018storm,
  title   = {Solar particle event storm shelter requirements for missions beyond low Earth orbit},
  author  = {Townsend, L. W. and Adams, J. H. and Blattnig, S. R. and Clowdsley, M. S. and Fry, D. J. and Jun, I. and McLeod, C. D. and Minow, J. I. and Moore, D. F. and Norbury, J. W. and Norman, R. B. and Reames, D. V. and Schwadron, N. A. and Semones, E. J. and Singleterry, R. C. and Slaba, T. C. and Werneth, C. M. and Xapsos, M. A.},
  journal = {Life Sciences in Space Research},
  volume  = {17},
  pages   = {32--39},
  year    = {2018},
  doi     = {10.1016/j.lssr.2018.02.002}
}
\bibliographystyle{unsrt}

\end{document}